\documentclass[sigconf]{acmart}

\usepackage{amsmath,amsthm,amssymb}
\usepackage[FIGTOPCAP]{subfigure}
\usepackage{multirow}
\setcopyright{rightsretained}
\newcommand{\mG}{\mathcal{G}}

\newcommand{\bbR}{\mathbb{R}}
\newcommand{\bu}{\mathbf{u}}
\newcommand{\bv}{\mathbf{v}}
\newcommand{\bw}{\mathbf{w}}
\newcommand{\bx}{\mathbf{x}}
\newcommand{\bb}{\mathbf{b}}

\newcommand{\new}[1]{{\color{black} #1}}

\newcommand{\ie}{\emph{i.e.},~}

\newcommand{\eg}{\emph{e.g.},~}

\newcommand{\ignore}[1]{}

\begin{document}
\title[]{An End-to-End Neighborhood-based Interaction Model for Knowledge-enhanced Recommendation}

\author{Yanru Qu$^{1*}$, Ting Bai$^{2,3*}$, Weinan Zhang$^{1}$, Jianyun Nie$^{4}$, Jian Tang$^{5,6,7}$} \thanks{$^*$ Equal contribution. This work was done when the first and the second authors were visiting Mila and Universit\'e de Montr\'eal.}
\affiliation{
\institution{$^{1}$Shanghai Jiao Tong University, $^{2}$Beijing University of Posts and Telecommunications, $^{3}$Renmin University of China \\ $^{4}$Universit\'e de Montr\'eal, $^{5}$Mila-Quebec Institute for Learning Algorithms, $^{6}$HEC Montr\'eal, $^{7}$CIFAR AI Research Chair}
}
\email{kevinqu@apex.sjtu.edu.cn,  baiting0317@gmail.com, wnzhang@sjtu.edu.cn}
\email{nie@iro.umontreal.ca, jian.tang@hec.ca}

\renewcommand{\shortauthors}{}

\begin{abstract}
\new{This paper studies graph-based recommendation, where an interaction graph is constructed from historical records and is leveraged to alleviate data sparsity and cold start problems. We reveal an early summarization problem in existing graph-based models, and propose Neighborhood Interaction (NI) model to capture each neighbor pair (between user-side and item-side) distinctively. NI model is more expressive and can capture more complicated structural patterns behind user-item interactions. To further enrich node connectivity and utilize high-order structural information, we incorporate extra knowledge graphs (KGs) and adopt graph neural networks (GNNs) in NI, called Knowledge-enhanced Neighborhood Interaction (KNI).
% To enrich the structural information, we also introduce graph neural networks (GNNs) and knowledge graphs (KGs) to NI, resulting an end-to-end model, namely Knowledge-enhanced Neighborhood Interaction (KNI). 
Compared with the state-of-the-art recommendation methods, \eg feature-based, meta path-based, and KG-based models, our KNI achieves superior performance in click-through rate prediction (1.1\%-8.4\% absolute AUC improvements) and outperforms by a wide margin in top-N recommendation on 4 real world datasets. 
% Our experiments on 4 real world datasets show that, compared with the state-of-the-art feature-based, meta path-based, and KG-based recommendation models, KNI achieves superior performance in click-through rate prediction (1.1\%-8.4\% absolute AUC improvements) and outperforms by a wide margin in top-N recommendation.
}
\end{abstract}


\copyrightyear{2019} 
\acmYear{2019} 
\acmConference[DLP-KDD'19]{1st International Workshop on Deep Learning Practice for High-Dimensional Sparse Data}{August 5, 2019}{Anchorage, AK, USA} 
\acmBooktitle{1st International Workshop on Deep Learning Practice for High-Dimensional Sparse Data (DLP-KDD'19), August 5, 2019, Anchorage, AK, USA}
\acmPrice{}
\acmDOI{10.1145/3326937.3341257}
\acmISBN{978-1-4503-6783-7/19/08}

\begin{CCSXML}
<ccs2012>
<concept>
<concept_id>10002951.10003317</concept_id>
<concept_desc>Information systems~Information retrieval</concept_desc>
<concept_significance>500</concept_significance>
</concept>
<concept>
<concept_id>10010147.10010257.10010293.10010294</concept_id>
<concept_desc>Computing methodologies~Neural networks</concept_desc>
<concept_significance>500</concept_significance>
</concept>
</ccs2012>
\end{CCSXML}

\ccsdesc[500]{Information systems~Information retrieval}
\ccsdesc[500]{Computing methodologies~Neural networks}

\keywords{Knowledge Graph, Knowledge-enhanced Recommendation, Neigh-borhood-based Interaction}

\maketitle

\section{Introduction}
Recommender systems have become increasingly important in various online services for helping users find the information they want. However, existing recommender systems are challenged by the problems of data sparsity and cold start, \ie most items receive only a few feedbacks (\eg ratings and clicks) or no feedbacks at all (\eg for new items).
\new{To tackle these problems, the existing approaches usually utilize side information to learn better user/item representations~\cite{koren2008factorization, he2018nais, kabbur2013fism}, which then facilitate the learning of user-item interactions, and finally promote the recommendation quality.
In many scenarios, knowledge graphs (KGs) can be used to provide general background knowledge as well as rich structural information~\cite{zhang2016collaborative, wang2018ripplenet, huang2018improving}.}

\new{Graph-based recommender systems build interaction graphs from historical feedbacks and side information, where the nodes can be users, items, or side information (\eg tag, genre), and two nodes are linked together based on relevance or co-occurrence. Recently, graph-based models are becoming more powerful with advanced graph neural networks. For example, graph convolution networks~\cite{van2017graph, ying2018graph} can integrate high-order neighborhood information in an end-to-end way, graph attention networks~\cite{wang2018ripplenet} can simulate user preferences on knowledge graphs.
Graph-based models are more expressive than traditional feature-based models, because they take the local structures (of the users, items, and relevant nodes) into consideration. 

However, due to an ``early summarization'' issue, the existing graph-based models cannot fully utilize the local structures:
these models usually compress user- and item-neighborhoods into single user/item embeddings before prediction. In this case, only two nodes and one edge are activated, yet other nodes and their connections are mixed and relayed. We consider the meticulous local structures are valuable, and a good system should be able to capture useful patterns, and filter out other noise. Here is an example. A system is recommending a film to a user, where the user has rated 5 stars for ``La La Land'' (Land) and ``Interstellar'' (Inter), and the film has 2 tags, ``romance'' and ``fiction''. We know that ``Land'' is a romance film, and ``Inter'' is a science fiction film. Thus the connections between (``Land'', ``romance'') and (``Inter'', ``fiction'') could be helpful for recommendation, meanwhile, the connection between (``Land'', ``fiction'') is nonsense and not expected, which is regarded as noise. We argue that the local structures are hidden and not fully utilized in previous graph-based methods.}

\new{To address the early summarization problem, we extend user-item interactions to their neighbors, and propose a unified Neighborhood Interaction (NI) model. More specifically, we propose a bi-attention network to make prediction on the local structures directly, instead of compressing them into user/item embeddings. We also utilize graph neural networks (GNNs) to integrate high-order neighborhood information, and introduce knowledge graphs to increase the local connectivity. The final model, called Knowledge-enhanced Neighborhood Interaction (KNI), is evaluated on 4 real-world datasets and compared with 8 feature-based, meta path-based, and graph-based models. Our experiments show that KNI outperforms state-of-the-art models, including Wide\&Deep, MCRec, PinSage and RippleNet, by 1.1\%-8.4\% of AUC in click-through rate prediction, and exceeds baseline models by a wide margin in top-N recommendation. We also provide a case study and statistical analysis to demonstrate our model.}

The rest of this paper is organized as follows: we first define the problem and introduce our KNI model in Section~\ref{sec:me}. And then we demonstrate the experiments and discuss the results in Section~\ref{sec:ex}. Related works are summarized in Section~\ref{sec:re}. Finally, Section~\ref{sec:co} concludes this paper.

\begin{figure*}
    \centering
    \includegraphics[width=\textwidth]{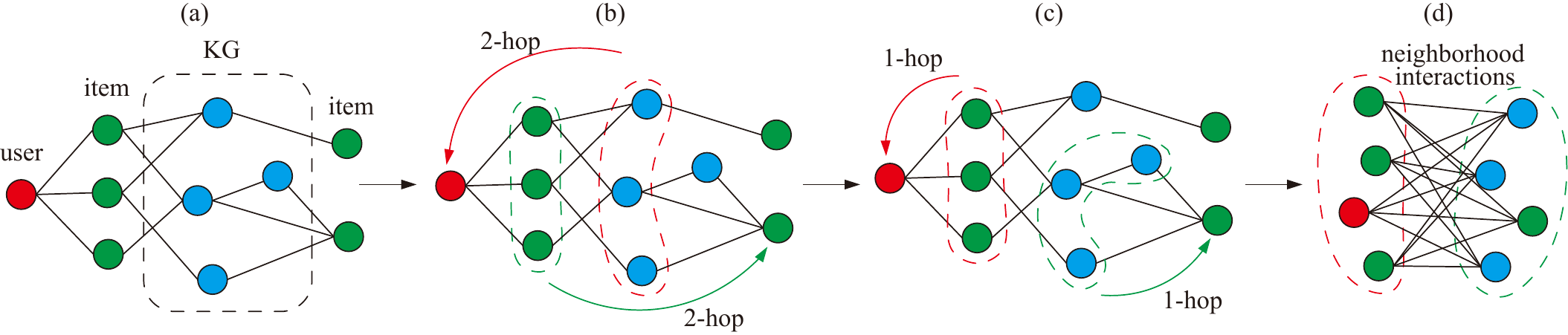}
    \caption{Model overview. \emph{Note}: Red circles denote users. Green circles denote rated or unseen items. Blue circles denote non-item entities. Dash circles denote user and item neighborhoods. In this example, a KIG is constructed at first, and then higher hop neighborhood information is aggregated into local neighbors. Finally, the user and item neighborhoods are collected to compute neighborhood interactions.} \label{fig:model}
\end{figure*}

\section{Knowledge-enhanced Neighborhood Interaction} \label{sec:me}
\new{In this section, we introduce graph-based recommendation and the early summarization issue at first, and define Neighborhood Interaction (NI). Then, we extend NI with graph neural networks (GNNs) and knowledge graphs (KGs). Finally, we present the overall framework of Knowledge-enhanced Neighborhood Interaction (KNI) model, shown in Fig.~\ref{fig:model}.}

\subsection{Neighborhood Interactions}\label{sec:ni}

% \new{Graph-based recommender systems build interaction graphs from historical responses. The graphs enrich the connectivity between users and items, and thus enhance the user-item interactions with structural information.}

\subsubsection{User-item Interaction Graph} The user-item interaction history can be represented by an interaction matrix, $Y \in \bbR^{|U| \times |V|}$, where $U = \{u_1, u_2, \dots, u_n\}$ is the set of users, $V = \{v_1, v_2, \dots, v_m\}$ the set of items.
An element $y_{u,v}$ indicates the feedback of user $u$ on item $v$. In this paper,
we assume that $y_{u,v}$ takes a binary value (which could be easily extended to other values). 
Regarding the positive responses in $Y$ as edges, we build the interaction graph for users and items, $\mG_{rec} = \{(u,c,v)|u \in U, v \in V, c = 1\}$. In $\mG_{rec}$, users' neighbors are items, and items' neighbors are users. 

\subsubsection{Graph-based Recommendation} \new{Most existing graph-based recommender systems utilize the graph structures via summarizing the neighborhood information into user/item representations. And then these models learn user-item interactions from the representations, which is usually formulated as inner product.} Denote $N_u$ and $N_v$ as user and item neighborhoods, $\bu$ and $\bv$ as their representations.
\begin{align}
\bu & = \text{agg}(N_u) \\
\bv & = \text{agg}(N_v) \\
\hat{y}_{u,v} & = \sigma(\langle \bu, \bv \rangle)
\end{align}
where $\text{agg}()$ is an aggregation function which maps a set of neighbor nodes into a single embedding vector, $\sigma()$ is the sigmoid function. For simplicity, we omit $\sigma$ in the following.

\new{The main difference of graph-based models is how to learn user/item representations from the graph structures, in another word, designing aggregation functions. The most popular and general $\text{agg}()$ include averaging, attention ~\cite{wang2018dkn,zhang2016collaborative,wang2018ripplenet}, etc.}
\begin{align}
\text{Average: } \bu & = \frac{1}{|N_u|} \sum_{i \in N_u} \bx_i \label{eq:avg_form} \\
\text{Attention: } \alpha_{u,i} & = \text{softmax}_i(\bw^\top [\bx_u, \bx_i] + b) \label{eq:atn_agg} \\
\bu & = \sum_{i \in N_u} \alpha_{u,i} \bx_i \label{eq:atn_form}
\end{align}
$\bx_u$ is the embedding vector of user $u$, $\bx_v$ is of item $v$, $\bx_i$ is of node $i$, and $\alpha_{u,i}$ is the user-side attention score produced by an attention network. \new{$\bw$ and $b$\footnote{In general, the bias term is canceled because of softmax. In \cite{velickovic2017graph}, the linear projection is followed by a LeakyReLU layer, therefore, the bias term is kept.} are attention network parameters. 
In this paper, we mainly employ the attention network structure in Eq.~\eqref{eq:atn_agg}, where $[,]$ means concatenation.}

\new{Most previous methods summarize user/item neighborhood information before learning their interactions, which compresses the local structures into only two nodes and one edge, yet other nodes and their connections are mixed and relayed. The behavior may restrict model from exploring the local structures and obstruct to distinguish useful patterns from noise. We call it the early summarization issue.}

\subsubsection{Neighbor-Neighbor Interaction} After expanding $\hat{y}_{u,v}$ (before taking $\sigma$), 
\begin{align}
\text{Average: } \hat{y}_{u,v} & = \langle \frac{1}{|N_u|} \sum_{i \in N_u} \bx_i, \frac{1}{|N_v|} \sum_{j \in N_v} \bx_j \rangle \\ 
& = \sum_{i \in N_u} \sum_{j \in N_v} \frac{1}{|N_u| |N_v|} \langle \bx_i, \bx_j \rangle \label{eq:avg}\\
\text{Attention: } \hat{y}_{u,v} & = \langle \sum_{i \in N_u} \alpha_{u,i} \bx_i, \sum_{j \in N_v} \alpha_{v,j} \bx_j \rangle \\
& = \sum_{i \in N_u} \sum_{j \in N_v} \alpha_{u,i} \alpha_{v,j} \langle \bx_i, \bx_j \rangle \label{eq:atn}
\end{align}
we find there exists a general form of Eq.~\eqref{eq:avg} and \eqref{eq:atn}
\begin{align}
& \hat{y} = \mathbf{A} \odot \mathbf{Z} \label{eq:form} \\
\text{s.t. } & \sum_{i,j} \mathbf{A}_{i,j} = 1 \text{ , } \mathbf{Z}_{i,j} = \langle \bx_i, \bx_j \rangle
\end{align}
where $\mathbf{A} \in \bbR^{|N_u| \times |N_v|}$ is a weight matrix summing up to 1, $\mathbf{Z} \in \bbR^{|N_u| \times |N_v|}$ is a matrix of inner product terms, and $\odot$ denotes the sum of element-wise product. According to this form, graph-based models generally learn two things: (i) exploring the general interactions of all node pairs (as $\mathbf{Z}$ does), (ii) assigning proper weights for different interactions (as $\mathbf{A}$ does). \new{We consider $\mathbf{Z}$ contains more global information because a node usually gets trained in different neighborhoods, yet $\mathbf{A}$ can better describe local structures, whose values indicate the confidence of certain connections being helpful. Therefore, this form, Eq.~\eqref{eq:form}, provides a tool to convert a graph structure to a prediction.}

We propose a bi-attention network to better utilize the neighborhood information, namely Neighborhood Interaction (NI).
\begin{align}
\alpha_{i,j} & = \text{softmax}_{i,j}(\bw^\top [\bx_u, \bx_i, \bx_v, \bx_j] + b) \label{eq:ni_atn}\\
\hat{y}_{u,v} & = \sum_{i \in N_u} \sum_{j \in N_v} \alpha_{i,j} \langle \bx_i, \bx_j \rangle \label{eq:ni}
\end{align}

Different from Eq.~\eqref{eq:atn_agg}, Eq.~\eqref{eq:ni_atn} takes both user- and item-side neighbors into consideration.
In Eq.~\eqref{eq:ni}, different interaction terms are weighted distinctively. From the discussion above, NI can better utilize the local structures for recommendation, therefore, NI is supposed to address the early summarization problem of graph-based models.

It is worth noting that the average form (Eq.~\eqref{eq:avg}) and attention form (Eq.~\eqref{eq:atn}) are special cases of NI. The average form sets the weight matrix as a constant $\mathbf{A} = 1/|N_u| |N_v|$, asnd the attention form approximates the weight matrix via 1st rank matrix decomposition $\mathbf{A} \sim \alpha_u \alpha_v^\top$, where $\alpha_u$ and $\alpha_v$ are column vectors of the user-/item-side attention scores (Eq.\eqref{eq:atn_agg}). 

Besides, we include the user and the item in their neighborhoods, \ie $u \in N_u$, $v \in N_v$, thus the interactions between user and item $(u, v)$, user and item neighbor $(u, j)$, user neighbor and item $(i, v)$, user neighbor and item neighbor $(i, j)$ are all considered for prediction.
The NI model is illustrated in Fig.~\ref{fig:model} (d), where the edges represent interactions among two neighborhoods.

\subsection{Integrating High-order Neighborhood Information}\label{sec:high-order}

\new{In last section, the user neighbors are ever-rated items, and the item neighbors are historical audience. The interaction graph $\mG_{rec}$ also contains high-order neighborhood information, for example, %a user is a 2-hop neighbor of another user if they have rated the same items, 
a film is a 2-hop neighbor of another film if they share the same tag. Introducing high-order neighborhood information has shown effective~\cite{van2017graph,ying2018graph,wang2017graphgan} in graph-based recommendation, thus we introduce graph convolution network (GCN)~\cite{kipf2016semi} and graph attention network (GAT)~\cite{velickovic2017graph} to encode high-order neighborhood information for NI model.}

\textbf{Graph convolution network} computes high-order node representations by stacking several graph convolution layers. Each graph convolution layer computes a node representation according to its nearest neighbors and itself (equivalent to a self loop in the graph). For a node $u$, a 2-layer GCN computes
\begin{align}
\bx_i^1 & = \sigma(\frac{1}{|N_i|} \sum_{j \in N_i} {\bw^1} \bx_j + \bb^1) \\
\bx_u^2 & = \sigma(\frac{1}{|N_u|} \sum_{i \in N_u} {\bw^2} \bx_i^1 + \bb^2) \label{eq:gcn}
\end{align}
where $\bx_j$ is the feature vector or initial embedding of node $j$, $\bx_i^1$ and $\bx_u^2$ are outputs of the 1st and 2nd graph convolution layers, $N_i$ and $N_u$ are neighborhoods of $i$ and $u$, and $\bw$ and $\bb$ are parameters to be learned. Successive graph convolution layers are separated by non-linear transformation $\sigma()$, which is usually ReLU.

\textbf{Graph attention network} is similar to GCN except that node embeddings are computed by multi-head self attention networks. For a node $u$, a 2-layer GAT computes (single head)
\begin{align}
\bx_i^1 & = \sigma(\sum_{j \in N_i} \alpha^1_{i,j} {\bw^1} \bx_j + \bb^1) \\
\bx_u^2 & = \sigma(\sum_{i \in N_u} \alpha^2_{u,i} {\bw^2} \bx_i^1 + \bb^2) \label{eq:gat}
\end{align}
where $\alpha^l_{i,j}$ is the attention score of node $j$ to node $i$, produced by the $l$-th layer attention network\footnote{GAT~\cite{velickovic2017graph} utilizes LeakyReLU transformation before softmax in its original form.}
\begin{align}
\alpha^l_{i,j} & = \frac{\exp(\text{LeakyReLU}({\bw_a^l}^\top [\bx_i^{l-1}, \bx_j^{l-1}] + b_a^l))}{\sum_{k \in N_i} \exp(\text{LeakyReLU}({\bw_a^l}^\top [\bx_i^{l-1}, \bx_k^{l-1}] + b_a^l))}
\end{align}
where $\bw_a^l$ and $b_a^l$ are parameters of the attention network, other notations are the same as GCN. Note that the above attention network structure is suggested in \cite{velickovic2017graph}.

For any target node $i$, we can generate node embeddings $\bx_i^l$ containing high-order neighborhood information with GCN or GAT~\cite{kipf2016semi,velickovic2017graph}. And $\{\bx_i^l, i \in \mG_{rec}\}$ can replace feature vectors or initial embeddings in NI model (Eq.~\eqref{eq:ni}), where graph network serves as an encoder. This process is demonstrated in Fig.~\ref{fig:model} (b) and (c). In (b) the 2-hop neighbors are propagated to 1-hop neighbors, and the 1-hop neighbors are concentrated to the central node.

\textbf{Neighbor Sampling} (NS)~\cite{hamilton2017inductive} is a sampling method to facilitate graph network computation on large graphs. 
The original graph networks, \eg GCN and GAT, traverse all neighbor nodes to generate a node embedding, which is time consuming and not tractable for a very large graph. 
NS proposes to sample a fixed number (\eg $K$) of neighbors for each node in forward computation. Combining GCN and NS as an example
\begin{align}
\tilde{\bx}^1_i & = \sigma(\frac{1}{K}\sum_{j \in \tilde{N}_i} {\bw^1} \bx_j + \bb^1)
\label{eq:ns}
\end{align}
where $\tilde{N}_i$ is drawn randomly from $N_i$, containing exactly $K$ elements. NS controls the number of high-order neighbors directly, thus restrains model's complexity.
There are other sampling methods, including random walk-based sampling \cite{ying2018graph}, importance sampling \cite{chen2018fastgcn}, etc. In this work, we mainly adopt NS.

\subsection{Integrating Knowledge Graphs}\label{sec:kig}

A knowledge graph consists of a large number of entity-relation-entity triples $\mG_{kg} = \{(h, r, t)| h, t \in E, r \in R\}$, where $E$ is the entity set, $R$ is the relation set.
Using the item set $V$ as initial queries, we can map items to corresponding entities in knowledge graph. 
Using the newly added entities as queries, we repeat the expansion several times and obtain knowledge-enhanced interaction graph (KIG),
$\mG = \mG_{rec} \cup \mG_{kg}$. 
The resulting KIG is shown in Fig.~\ref{fig:model} (a). In KIG, the users' and items' neighbors are extended to non-item entities, \eg a movie star. We can recklessly replace $\mG_{rec}$ with $\mG$ without modifying NI.

\subsection{Model Overview}\label{sec:obj}
The training objective is log loss
\begin{align}
\mathcal{L}(Y, \hat{Y}) = - \sum_{y_{u,v} = 1} \log(\hat{y}_{u,v}) - \sum_{y_{u,v} = 0} \log(1 - \hat{y}_{u,v}) + \lambda \Vert \theta \Vert_2^2 \label{eq:loss}
\end{align}
where $\lambda \Vert \theta \Vert_2^2$ is the L2 regularization term to control overfitting.

We then revisit the whole framework of KNI as shown in Fig.~\ref{fig:model}. (a): We first build knowledge-enhanced interaction graph (KIG) with user feedbacks and knowledge graphs. (b) and (c): We then apply graph neural networks (GNNs) to propagate high-order neighborhood information to user/item neighbors. (d): The user and item neighborhoods are collected and fed to Neighborhood Interactions (NI). The framework is trained end-to-end with the loss term presented above.

\section{Experiments} \label{sec:ex}

\begin{table}
\centering
\caption{Statistics for the expanded datasets. \emph{Note}: ``entities'' contain both items and non-item entities.}
\resizebox{\columnwidth}{!}{
\begin{tabular}{c|cccc} \hline \label{tab:st}
 Datasets & C-Book & Movie-1M & A-Book & Movie-20M \\ \hline\hline
\# users & 17,860 & 6,036 & 78,809 & 59,296 \\
\# items & 14,967 & 2,445 & 32,389 & 11,895 \\
\# interactions & 139,746 & 753,772 & 1,181,684 & 9,104,038 \\
\# entities & 77,881 & 182,011 & 265,478 & 64,067 \\
\# relations & 10 & 12 & 22 & 38 \\
\# triples & 71,628 & 923,718 & 1,551,554 & 1,195,391 \\ \hline
\end{tabular}}
\end{table}

\subsection{Datasets}

We combine 4 recommendation datasets with 2 public knowledge graphs in our experiments. The datasets and experiment code are publicly available\footnote{https://github.com/Atomu2014/KNI} for reproducibility and further study.

The first two smaller datasets are released by \cite{wang2018ripplenet}.

\begin{itemize}
\item \textbf{C-Book} combines Book Crossing\footnote{http://www2.informatik.uni-freiburg.de/~cziegler/BX/} and Microsoft Satori\footnote{https://searchengineland.com/library/bing/bing-satori}.
\item \textbf{Movie-1M} combines MovieLens\footnote{https://grouplens.org/datasets/movielens/}-1M and and Microsoft Satori.
\end{itemize}

We follow the procedures of \cite{wang2018ripplenet} to process the other two larger datasets, which are then linked to Freebase \cite{bollacker2008freebase}. The linkages are studied and provided by KB4Rec \cite{zhao2018kb4rec}. Note that another dataset LFM in KB4Rec is not included in our experiments, because it follows a quite different scheme from the others and does not contain any rating or click information.

\begin{itemize}
\item \textbf{A-Book} combines Amazon Book\footnote{http://jmcauley.ucsd.edu/data/amazon/} and Freebase. Amazon Book \cite{he2016ups}
contains over 22.5 million ratings (ranging from 1 to 5) collected from 8 million users and 2.3 million items.
\item \textbf{Movie-20M} combines MovieLens-20M and Freebase. Movie-Lens-20M contains ratings (ranging from 1 to 5) collected from the MovieLens website.
\end{itemize}

% The volumes of the above datasets follow: C-Book $<$ Movie-1M $<$ A-Book $<$ Movie-20M.
A-Book and Movie-20M are processed as follows.
Since A-Book and Movie-20M are originally in rating format, we convert ratings to binary feedbacks: 4 and 5 stars are converted to positive feedbacks (denoted by ``1'') and the other ratings to negative feedbacks.
For each user, we sample the same amount of negative samples (denoted by ``0'') as their positive samples from unseen items.
%Thus positive samples compose around 50\% of each dataset.
We also drop low-frequency users and items.
The threshold is 5 for A-Book and  20 for Movie-20M. 

After the datasets are processed, we split each dataset into training/validation/test sets at 6:2:2.
Then we map the items of training set to corresponding entities in Freebase, with the help of KB4Rec ~\cite{zhao2018kb4rec}.
For each dataset, we use the linked items as initial queries to find related non-item entities.
These entities are added to KIG and used for further expansion.
We repeat this process 4 times to ensure sufficient knowledge is included in the final dataset.
We also remove entities appearing less than 5 times on A-Book (the threshold is 20 for Movie-20M), and relations appearing less than 5000 times (same for Movie-20M).
The basic statistics of the 4 datasets are presented in Table~\ref{tab:st}.

\subsection{Compared Models}

We compare NI (without knowledge graph) and KNI (with knowledge graph) with 2 feature-based, 2 meta path-based, and 4 graph-based models. For fair comparison, we pre-train TransR models and extract structural features for feature-based models. \new{Note that the main difference between NI models and baseline models is, NI models make prediction from graph structures directly, while others compress the structural information into two node embeddings. From another perspective, NI models further take the interactions between neighbors into consideration.}

\textbf{libFM} \cite{rendle2010factorization} is a widely used feature-based model, and is well known for modeling feature interactions. In our experiments, we concatenate the user ID, item ID, and the average embedding of related entities learned from TransR \cite{lin2015learning} as the input to libFM.

\textbf{Wide\&Deep} \cite{cheng2016wide} is another feature-based model, which takes  advantages of both shallow models and deep models and achieves state-of-the-art recommendation results. We provide the same input as libFM to Wide\&Deep.

\textbf{PER} \cite{yu2014personalized} is a meta path-based model, which builds heterogeneous information network (HIN) on side information, and extracts meta path-based features from HIN. 
In our experiments, we use all item-attribute-item relations as meta paths. 

\textbf{MCRec}~\cite{hu2018leveraging} is a co-attentive model built on HIN. MCRec learns context representations from meta-paths, and is a state-of-the-art recommendation model. Besides, their code is released\footnote{https://github.com/librahu/MCRec}.

\textbf{CKE}~\cite{zhang2016collaborative} proposes a general framework to jointly learn structural/textual/visual embeddings from knowledge graph, texts and images for collaborative recommendation. We adopt the structural embedding and recommendation components of CKE.

\textbf{DKN}~\cite{wang2018dkn} is another knowledge graph-based recommendation model. In our experiments, we use pre-trained TransR embeddings as the input for DKN, with code\footnote{https://github.com/hwwang55/DKN}.

\textbf{PinSage} \cite{ying2018graph} uses GCN for web-scale recommendation. 
In our experiments, we use PinSage as a representative GCN approach and explore different network structures and sampling methods on PinSage.

\textbf{RippleNet}~\cite{wang2018ripplenet} is a state-of-the-art knowledge graph-based recommendation model. RippleNet uses attention networks to simulate user preferences on KG. In our experiments, we use RippleNet as a representative  GAT approach, with code\footnote{https://github.com/hwwang55/RippleNet}.

It is worth noting that, Wide\&Deep, MCRec, PinSage and RippleNet are recently proposed state-of-the-art models.

\begin{figure}
    \subfigure{\centering \includegraphics[width=0.23\textwidth]{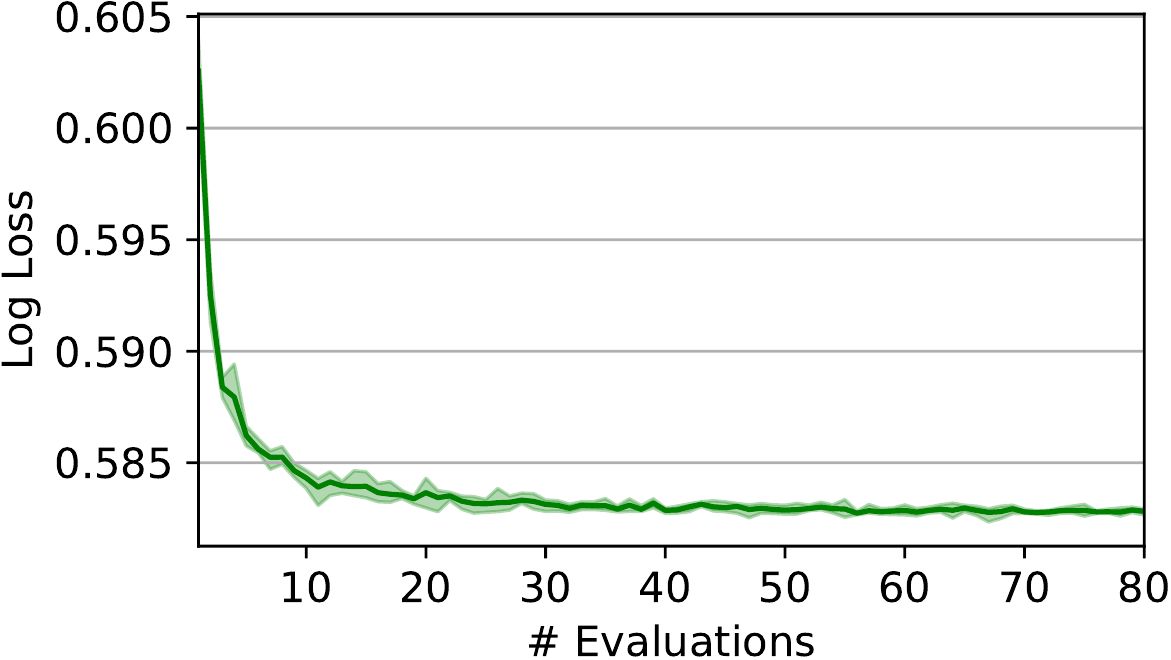}}
    \subfigure{\centering \includegraphics[width=0.23\textwidth]{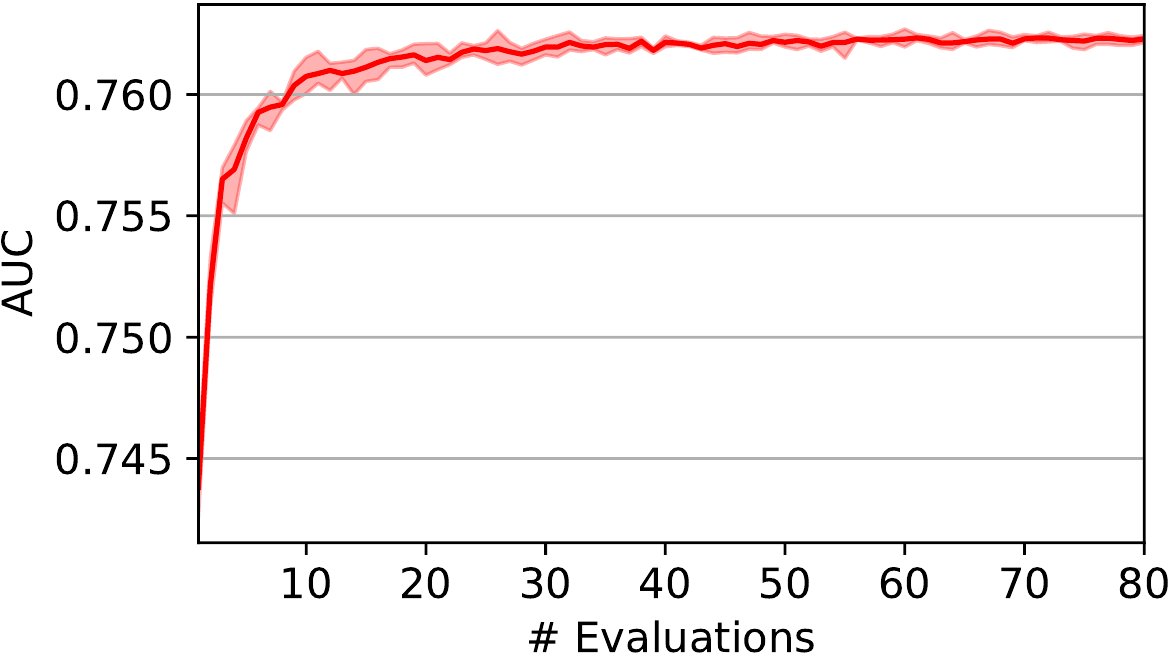}}
    \caption{Evaluation stabilizes after sufficient evaluations.} \label{fig:n_eval}
\end{figure}

\subsection{Experiment Setup and Evaluation}

We evaluate these models on 2 tasks, click-through rate (CTR) prediction and top-N recommendation.
For CTR prediction, we use the metrics Area Under Curve (AUC) and Accuracy (ACC), which are widely used in binary classification problems.
For top-N recommendation, we use the best models obtained in CTR prediction to generate top-N items, which are compared with the test set to compute Precision@K, Recall@K, and F1@K.
We repeat each experiment 5 times and report the average scores.

General hyper-parameters include learning rate, embedding size, regularization, etc.
Graph-based models, including PinSage, RippleNet and our models, are trained with graph network modules.
For these models, 2 hyper-parameters are critical, \ie the hop number and the sampling method.

A larger hop number indicates a larger neighborhood.
In the graph construction stage, we expand the items 4 times on Freebase, thus an item needs 4 steps to visit certain neighbors.
For graph-based models, we tune the hop number from 1 to 4.
Sampling methods are mainly introduced to speed up training on large graphs, and sometimes influence model convergence and performance.
We tune neighbor sampling (NS) and random walk-based sampling in experiments. 

We then apply grid search on embedding dimension, learning rate, l2 regularization, etc.,  for all the compared models.
The hyper-parameters are chosen according to the AUC scores on validation sets, and the parameter settings are explained in Section~\ref{sec:param}.

We repeat evaluation several times and use the average scores to compute the metrics. We perform an empirical experiments to determine the number of repetitions, shown in Fig.~\ref{fig:n_eval}. According to the figure, we conclude the prediction becomes stable after sufficient evaluations.
In the following experiments, we fix this number to 40.

\subsection{Experiment Results}

\begin{table*}
\centering
\caption{The results of CTR prediction. \emph{Note}: ``*'' indicates the statistically significant improvements over the best baseline, with $p$-value smaller than $10^{-6}$ in two-sided $t$-test.}\label{tab:ctr}
\begin{tabular}{c|cc|cc|cc|cc} \hline
\multirow{2}{*}{Model} & \multicolumn{2}{|c|}{C-Book} & \multicolumn{2}{|c|}{Movie-1M} & \multicolumn{2}{|c|}{A-Book} & \multicolumn{2}{|c}{Movie-20M} \\ \cline{2-9}
& AUC & ACC & AUC & ACC & AUC & ACC & AUC & ACC \\ \hline\hline
libFM & 0.6850 & 0.6390 & 0.8920 & 0.8120 &	0.8300 & 0.7597 & 0.9481 & 0.8805\\
Wide\&Deep & 0.7110 & 0.6230 & 0.9030 &	0.8220 & 0.8401 & 0.7684 & 0.9507 &	0.8831 \\ \hline
PER & 0.6230 & 0.5880 & 0.7120 & 0.6670 & 0.7392 & 0.6939 & 0.8161 & 0.7327 \\
% SVD++ & 0.7026 & 0.6295 & 0.91424 & 0.8237 & 0.8646 & 0.7792 & 0.9591 & 0.8953 \\
MCRec & 0.7250 & 0.6707 & 0.9127 & 0.8331 & 0.8708 & 0.7930 & 0.9558 & 0.8872 \\ \hline
%CKE & 0.6740 & 0.6350 & 0.7960 & 0.7390 & 0.8572 & 0.7839 & 0.9574 & 0.8940 \\
CKE & 0.6760 & 0.6422 & 0.8974 & 0.8171 & 0.8572 & 0.7839 & 0.9574 & 0.8940 \\
DKN & 0.6488 & 0.6333 & 0.8835 & 0.8070 & 0.8455 & 0.7679 & 0.9473 & 0.8787 \\ 
PinSage & 0.7102 & 0.6477 & 0.9213 & 0.8443 & 0.8634 & 0.7804 & 0.9597 & 0.8960 \\
RippleNet & 0.7290 & 0.6630 & 0.9210 & 0.8440 & 0.8736 & 0.7975 & 0.9579 & 0.8942\\ \hline
% KIG + Max & 0.7199 & 0.6731 & 0.9256 & 0.8441 & 0.8260 & 0.7431 & 0.9409 & 0.8683 \\
% KIG + GC & 0.7626 & 0.7003 & 0.9401 & 0.8672 & 0.9229 & 0.8457 & 0.9658 & 0.9054\\
% KIG + GA & 0.7679 & 0.7023 & 0.9415 & 0.8703 & 0.9236 & 0.8468 & 0.9685 & 0.9093 \\
NI & 0.7468 & 0.6796 & 0.9401 & 0.8679 & 0.9160 & 0.8362 & 0.9693 & 0.9110 \\
KNI & \textbf{0.7723}* & \textbf{0.7063}* & \textbf{0.9449}* & \textbf{0.8721}* & \textbf{0.9238}* & \textbf{0.8472}* & \textbf{0.9704}* & \textbf{0.9120}* \\ \hline
\end{tabular}
\end{table*}

\begin{figure*}
    \centering
    \includegraphics[width=0.9\textwidth]{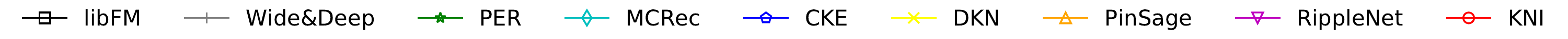}
    \subfigure[Precision@K]{
        \centering
        \includegraphics[width=0.32\textwidth]{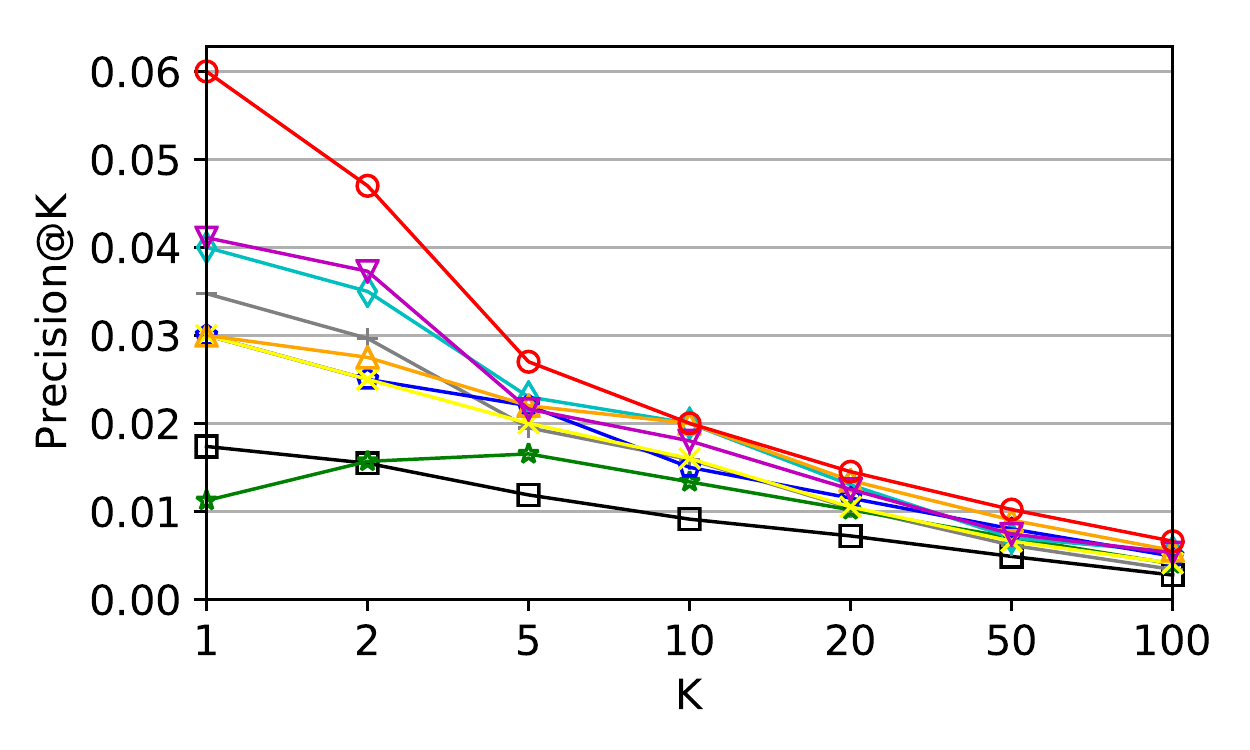}
    }
    \subfigure[Recall@K]{
        \centering
        \includegraphics[width=0.32\textwidth]{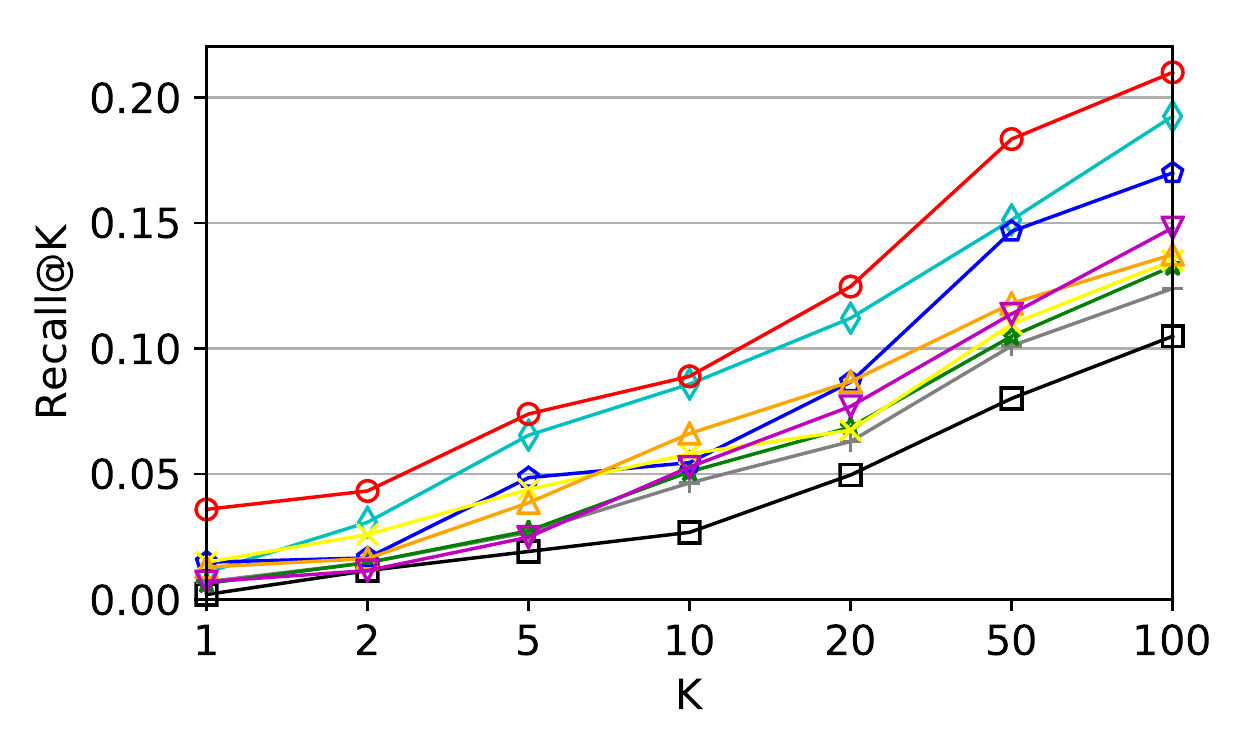}
    }
    \subfigure[F1@K]{
        \centering
        \includegraphics[width=0.32\textwidth]{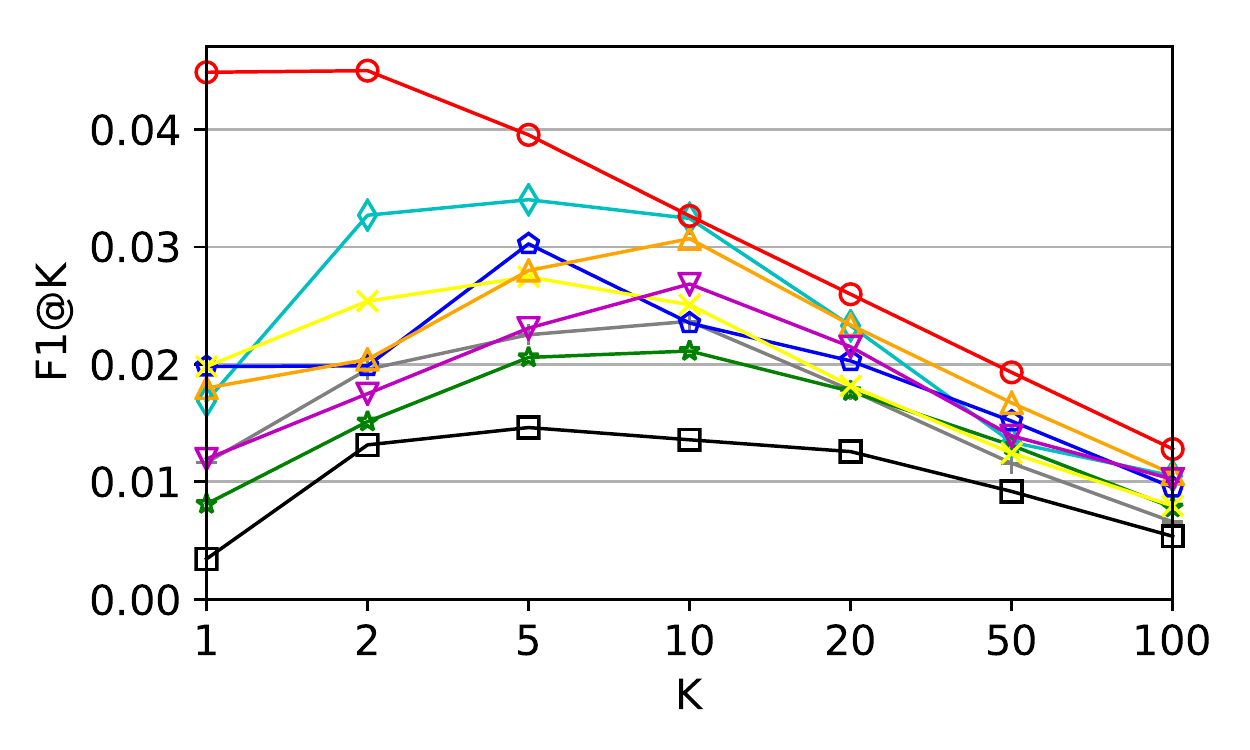}
    }\vspace{-10pt}
    \caption{Top-N recommendation results for C-Book.} \label{fig:topk_bc} \vspace{10pt}
    \subfigure[Precision@K]{
        \centering
        \includegraphics[width=0.32\textwidth]{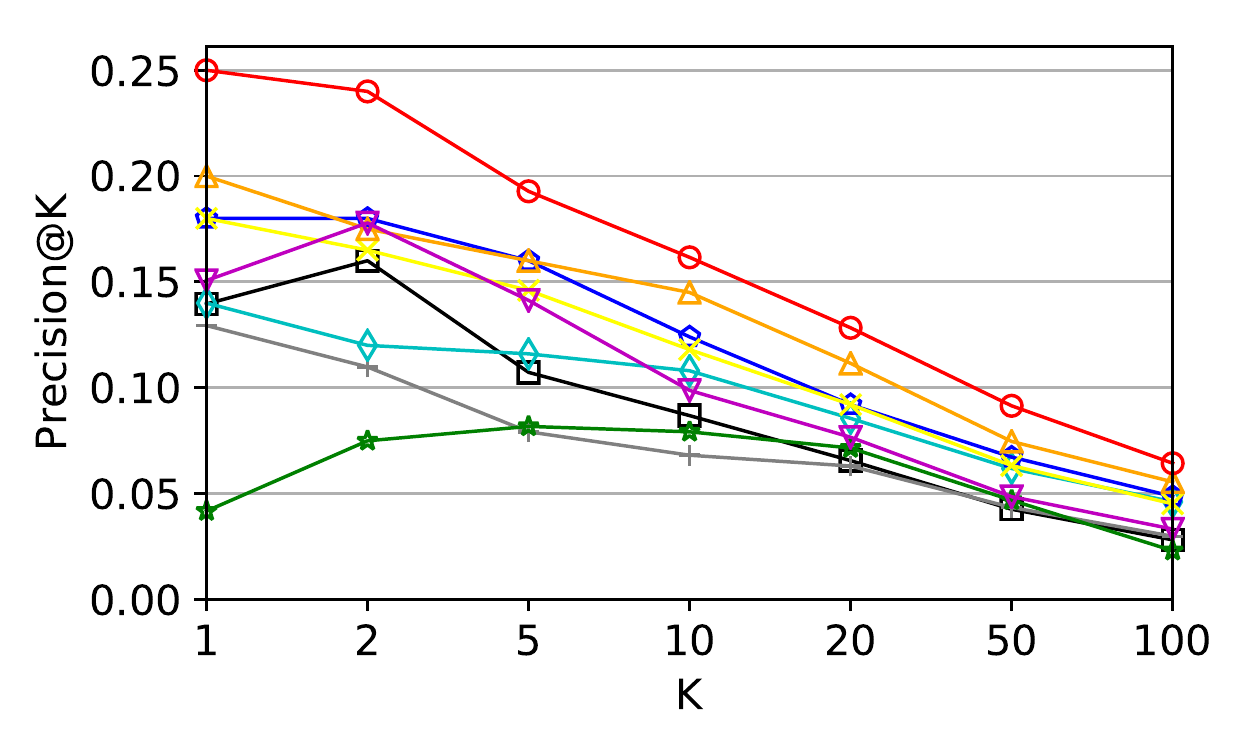}
    }
    \subfigure[Recall@K]{
        \centering
        \includegraphics[width=0.32\textwidth]{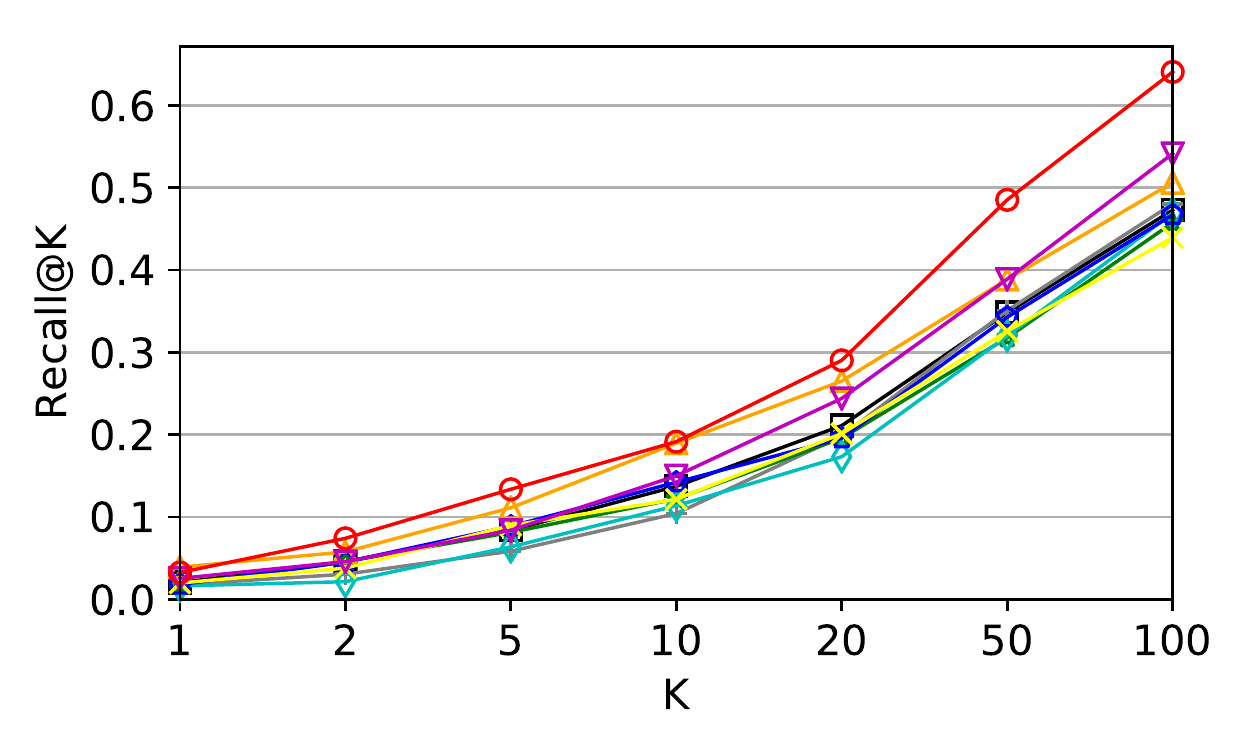}
    }
    \subfigure[F1@K]{
        \centering
        \includegraphics[width=0.32\textwidth]{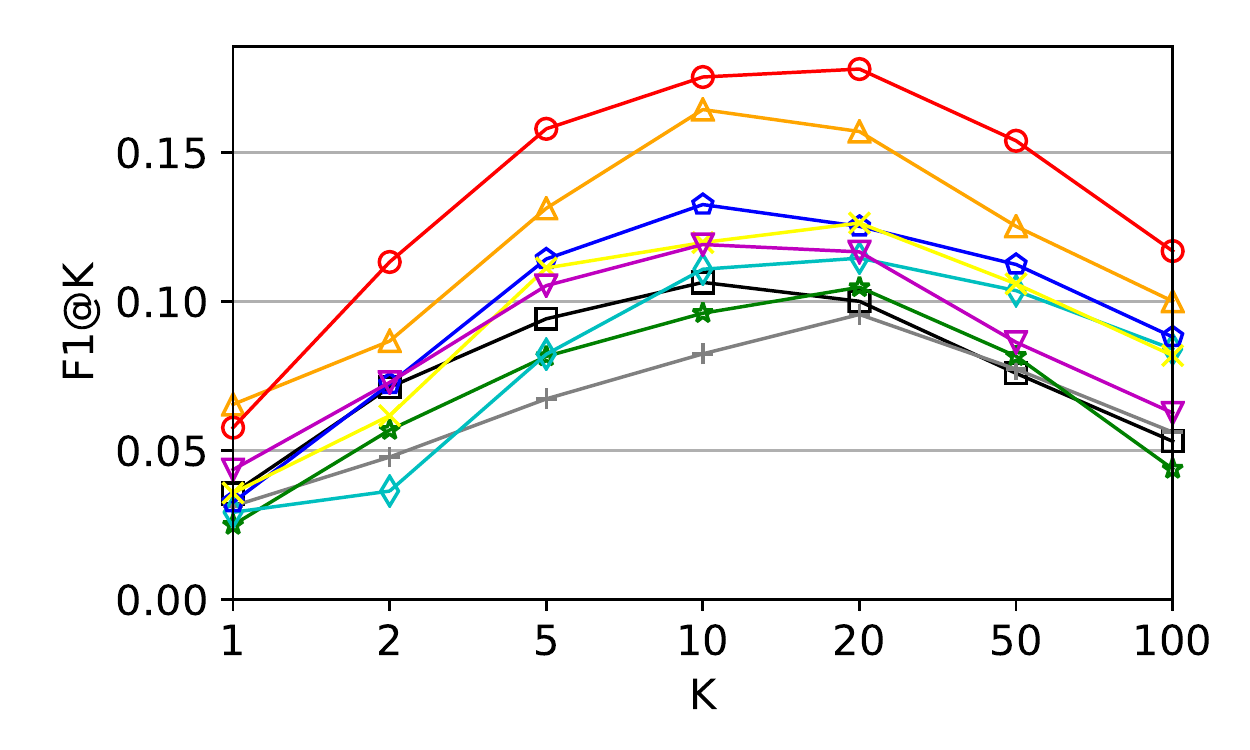}
    } \vspace{-10pt}
    \caption{Top-N recommendation results for Movie-1M.} \label{fig:topk_m1} \vspace{10pt}
    \subfigure[Precision@K]{
        \centering
        \includegraphics[width=0.32\textwidth]{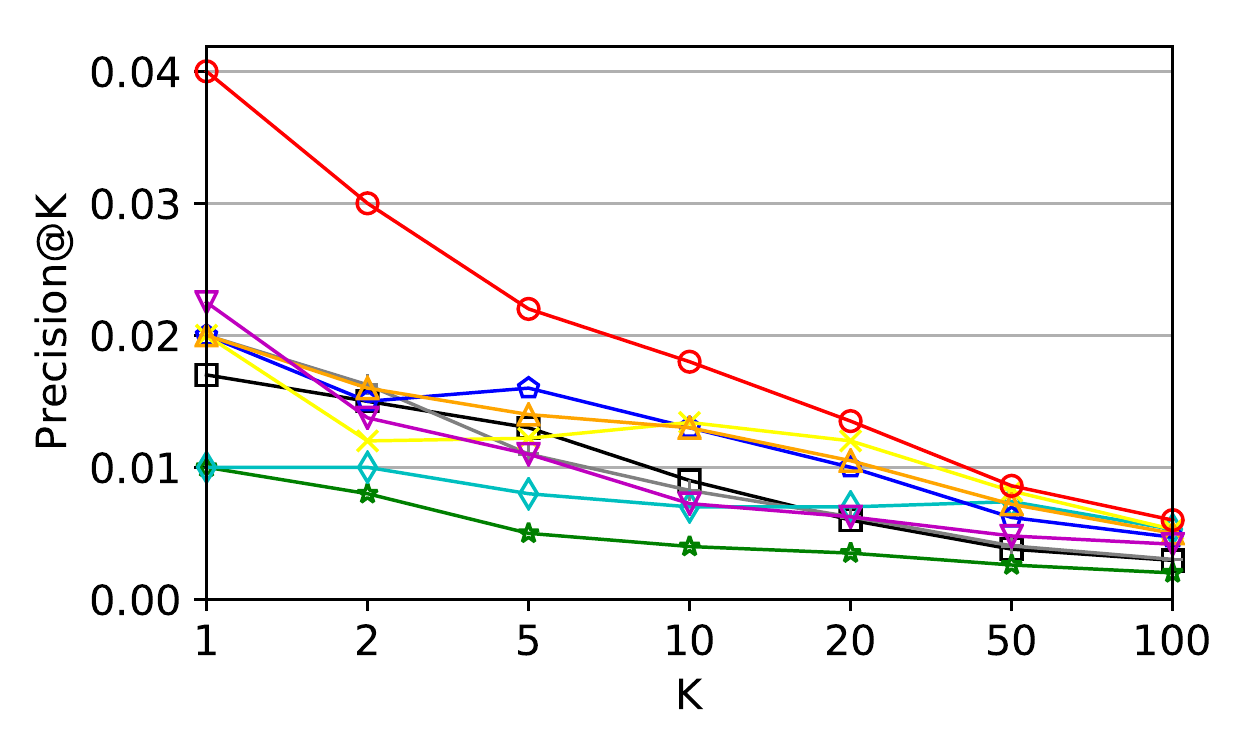}
    }
    \subfigure[Recall@K]{
        \centering
        \includegraphics[width=0.32\textwidth]{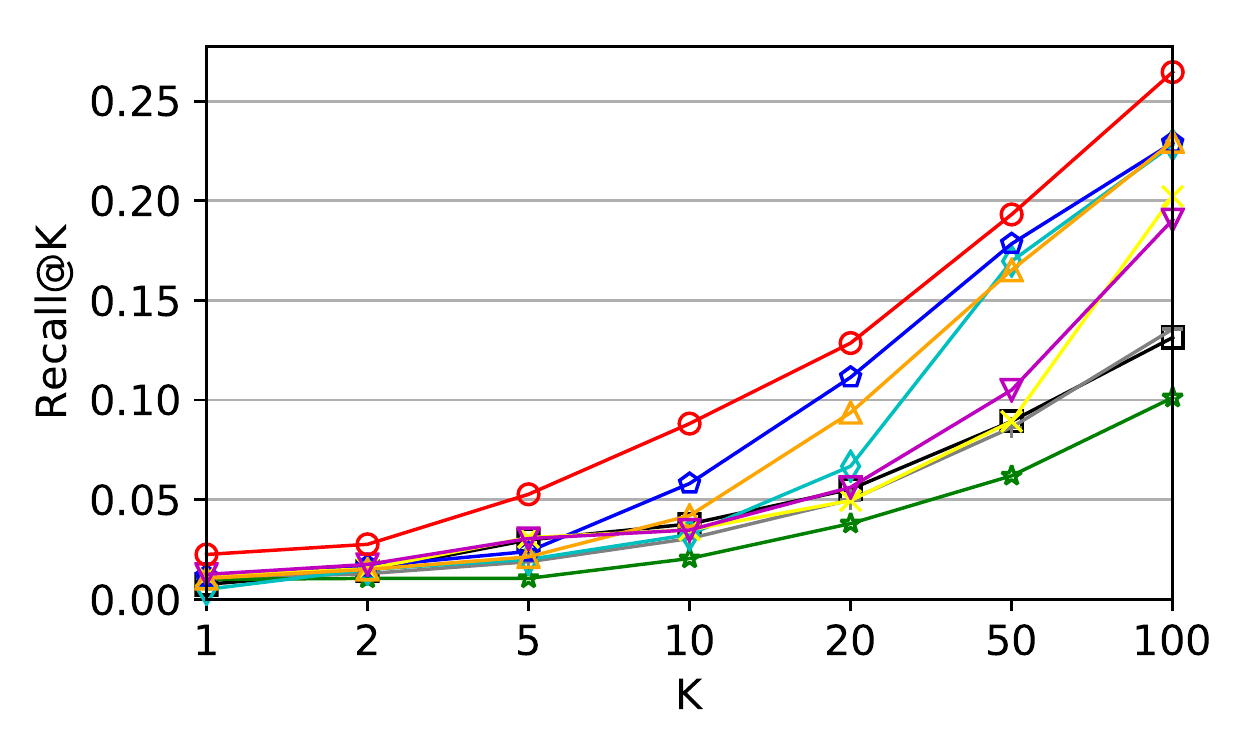}
    }
    \subfigure[F1@K]{
        \centering
        \includegraphics[width=0.32\textwidth]{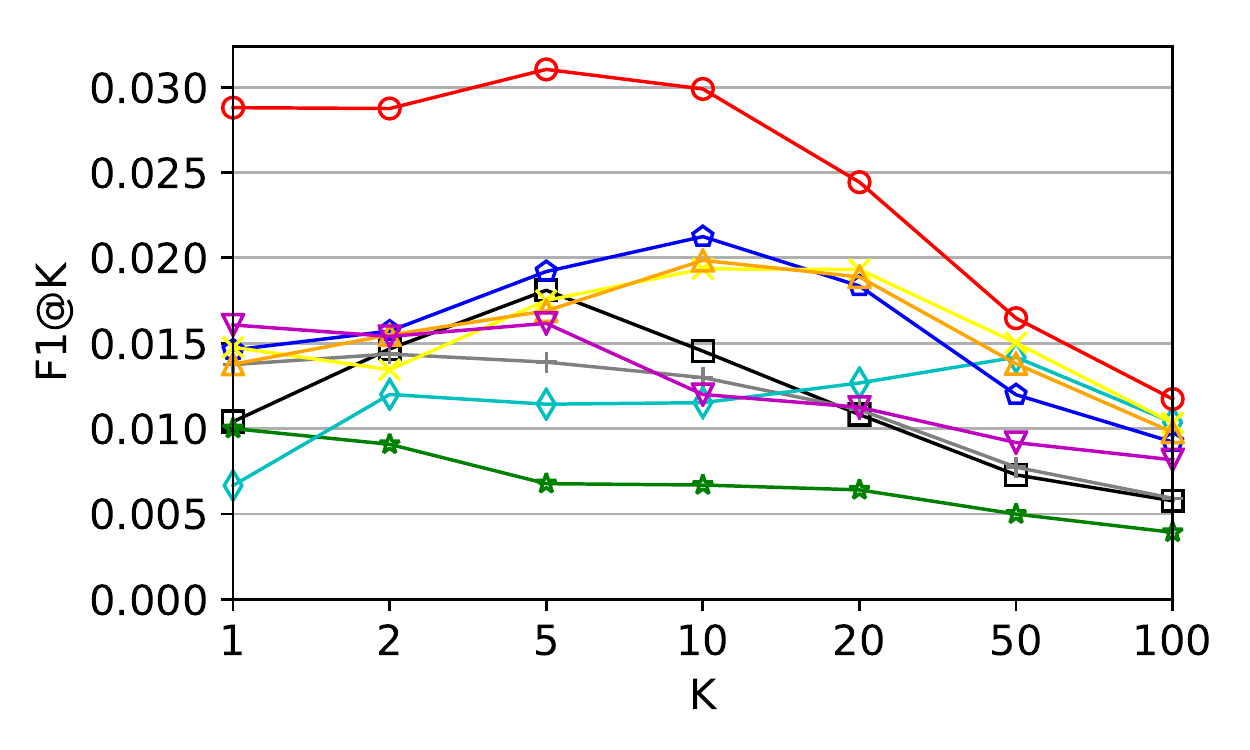}
    } \vspace{-10pt}
    \caption{Top-N recommendation results for A-Book.} \label{fig:topk_ab} \vspace{10pt}
    \subfigure[Precision@K]{
        \centering
        \includegraphics[width=0.32\textwidth]{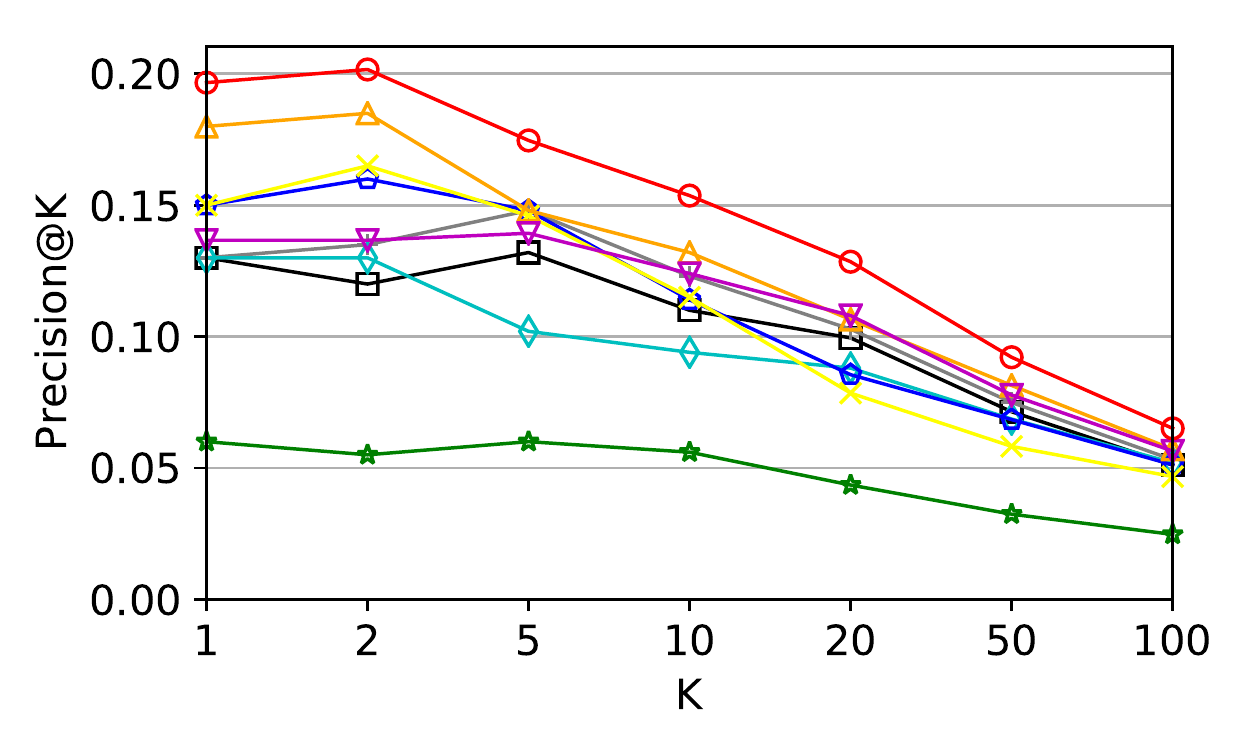}
    }
    \subfigure[Recall@K]{
        \centering
        \includegraphics[width=0.32\textwidth]{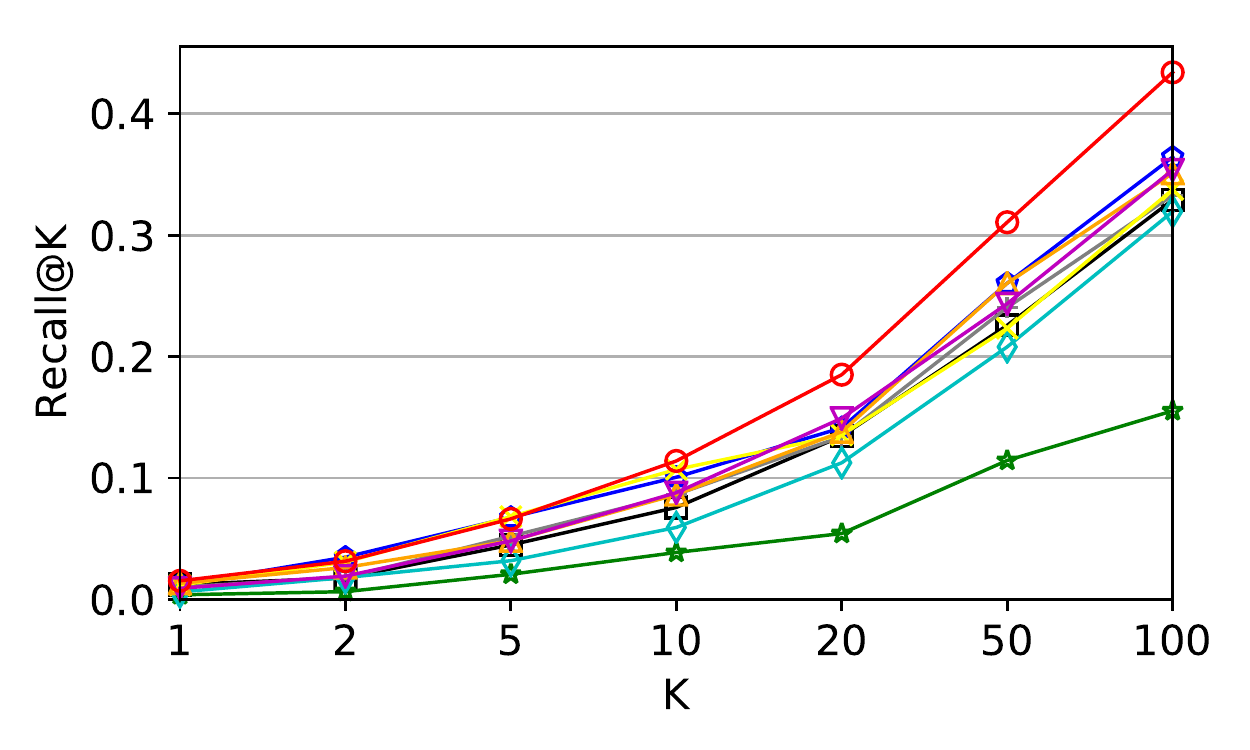}
    }
    \subfigure[F1@K]{
        \centering
        \includegraphics[width=0.32\textwidth]{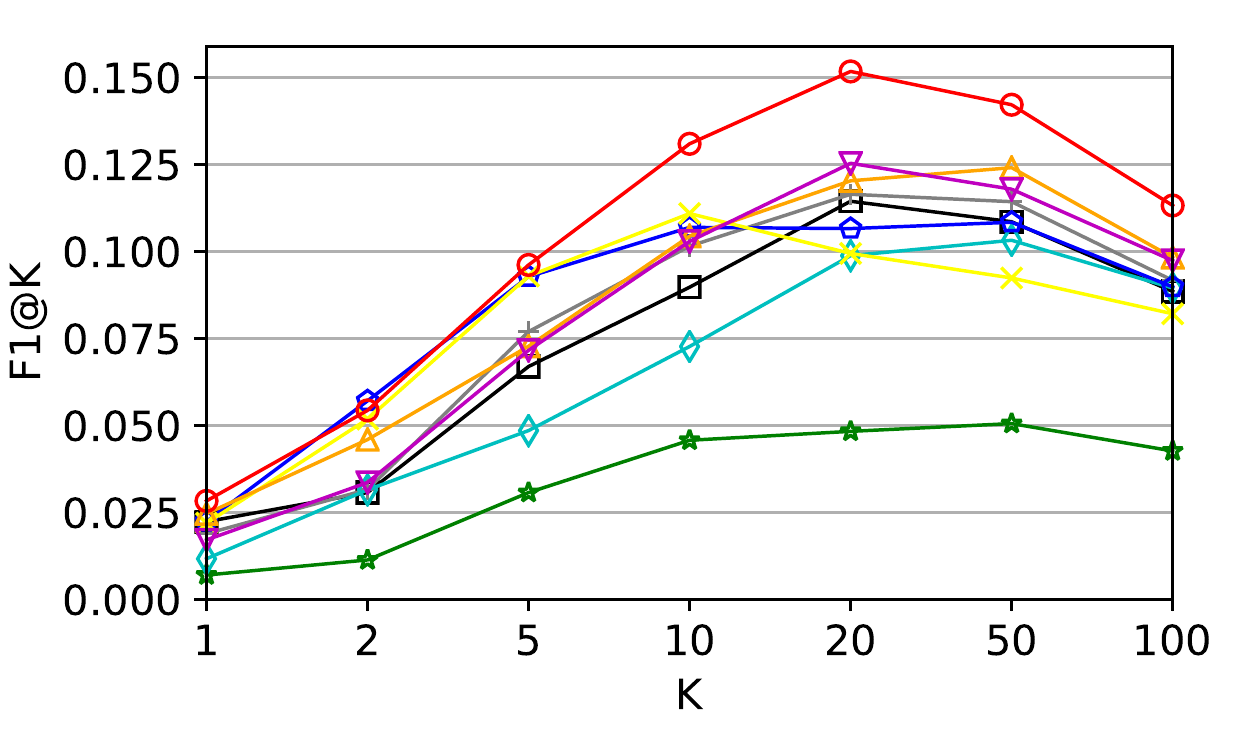}
    } \vspace{-10pt}
    \caption{Top-N recommendation results for Movie-20M.} \label{fig:topk_m20}
\end{figure*}

In this section, we present and analyze the evaluation results of CTR prediction (Table~\ref{tab:ctr}) and top-N recommendation (Fig.~\ref{fig:topk_bc}, \ref{fig:topk_m1}, \ref{fig:topk_ab}, \ref{fig:topk_m20}).
From Table~\ref{tab:ctr} we can observe:

(i) Meta path-based and graph-based models outperform feature-based models. MCRec, PinSage and RippleNet outperform the other baseline models. %\new{One possible reason is that meta path-based and graph-based models further utilize structural information.}

(ii) Meta-path design requires much human expertise, and is not end-to-end. 
Even though MCRec achieves competitive results with RippleNet, it requires more efforts to manually design and pre-process meta-paths. This restricts the application of meta path-based models on large graphs and scenarios with complex schema.

(iii) High-order neighborhood information contains much more noise. 
We increase the hop numbers of different models from 1 to 4, and find performance usually decreases with 3- or 4-hops.
We attribute this problem to the noise brought by the huge amounts of high-order neighbors (Table~\ref{tab:sp}). 

(iv) NI shows significant improvements over baseline models. To our surprise, NI outperforms PinSage and RippleNet even without knowledge graphs. \new{This means the local neighborhood structures are more valuable than high-order neighbors. We also observe that high-order neighborhoods increase dramatically.}

(v) Integrating knowledge graphs, KNI obtains even better results than NI. \new{Compared with Wide\&Deep, MCRerc, PinSage, and RippleNet, KNI achieves 1.1\%-8.4\% AUC improvements on 4 datasets.} %With these results, we confirm that KNI is an effective end-to-end knowledge-enhanced recommender system.

(vi) \new{From the data perspective, book datasets are more sparse ($>$ 99.9\%) than movie datasets, according to Table~\ref{tab:sp}. However, KNI achieves better improvements on the 2 book datasets (4\%-5\% AUC improvements over best baselines) than the 2 movie datasets (1\%-2\% AUC improvements). This means KNI can better solve data sparsity.}

\begin{table}
    \centering
    \caption{Data sparsity statistics and AUC improvements. \emph{Note}: The n-hop columns represent the number of n-hop neighbors. The sparsity is calculated as \# missing edges / \# node pairs. The improvements are absolute AUC gains of KNI compared with best baselines.} \label{tab:sp}
    \small
    \begin{tabular}{c|ccc|cc} \hline
    Datasets& 1-hop & 2-hop & 3-hop & Sparsity & Improvement \\ \hline\hline
    C-Book & 1 & 58 & 40 & 99.97\% & 4.33\% \\
    Movie-1M & 14 & 42,227 & 35,534 & 97.45\% & 2.36\% \\
    A-Book & 5 & 17,027 & 49,419 & 99.98\% & 5.02\%\\
    Movie-20M & 17 & 40,547 & 14,966 & 99.35\% & 1.07\% \\ \hline
    \end{tabular}
\end{table}

For the top-N recommendation task, we compare KNI with baseline models.
From Fig.~\ref{fig:topk_bc}, \ref{fig:topk_m1}, \ref{fig:topk_ab}, and \ref{fig:topk_m20} we can observe:

(i) The top-N recommendation results are consistent with CTR prediction. Meta path-based and graph-based models perform better than feature-based models. KNI performs the best.
    
(ii) On the two book datasets, KNI performs much better than baselines when K is small, especially in top-1 recommendation. This indicates that KNI captures user preference very well. On the 2 movie datasets, KNI outperforms state-of-the-art baseline models by a wide margin.

\subsection{Parameter Settings}\label{sec:param}

For hop number, we tune RippleNet following \cite{wang2018ripplenet}, and find RippleNet performs best with hop=3 (C-Book), hop=2 (Movie-1M), hop=1 (A-Book and Movie-20M).
For PinSage and our models, we find 1-hop is good enough. 
After an analysis of the datasets, we found that the main reason for this problem is the explosive increase of high-order neighbors.
From Table~\ref{tab:sp} we can see that the average neighborhood size increases dramatically when it goes from 1-hop to 2-hop, especially on the 2 movie datasets.
This may be caused by some high degree nodes in the knowledge graph. 
The noise brought by the high-order neighbors increases training difficulties.
Similar results can be found in many other studies. For example,
\cite{wang2018ripplenet} shows that larger hop numbers may decrease model performance. 
In \cite{van2017graph}, the author claims that 1 layer GCN performs the best.

As for the sampling method, we tune NS and random walk-based sampling on PinSage.
We find that random walk-based sampling does not always produce better results than NS, besides, random walk-based sampling requires more time.
Thus we only apply NS on the other models.
The number of neighbors to be sampled is tuned from \{4, 8, 16, 32, 64, 128\} (128 is not applicable on A-Book and Movie-20M due to memory constraints), and we find 4 (C-Book), 32 (Movie-1M), 8 (A-Book), and 32 (Movie-20M) perform slightly better.
We also test the training speed of RippleNet and KNI. When fixing the maximum neighbor size to be 32, KNI with NS could be 5.6-8.6 times faster than RippleNet to train one iteration in the same GPU environment, shown in Table~\ref{tab:tr}. This result confirms that the model complexity of KNI (Section~\ref{sec:obj}) could be well controlled through sampling and parallelization.

We perform grid search on the embedding dimension, learning rate and l2 regularization for each model, and we find that the embedding dimension 128 is the best of \{4, 8, 16, 32, 64, 128\} (we do not try higher dimensions considering the memory size), and the learning rate $10^{-3}$ is generally better than $\{10^{-4}, 2*10^{-4}, 5*10^{-4}, 2*10^{-3}, 5*10^{-3}, 10^{-2}\}$ (different models vary slightly), and we set the L2 regularization differently on different datasets: $10^{-5}$ (C-Book), $10^{-7}$ (Movie-1M), $10^{-7}$ (A-Book), $10^{-8}$ (Movie-20M). For other hyper-parameters provided by open-source softwares, we tune them carefully in the grid search.

\begin{table}
    \centering
    \caption{Training time of RippleNet and KNI.}
    \label{tab:tr}
    \begin{tabular}{c|cccc} \hline
        Models & C-Book & Movie-1M & A-Book & Movie-20M \\ \hline \hline
        RippleNet & 17.75s & 66.85s & 120.38s & 937.92s \\
        KNI & 2.05s & 11.58s & 21.52s & 166.72s \\ \hline
    \end{tabular}
\end{table}

\subsection{Case Study}\label{sec:cs}
\begin{figure}
    \subfigure[Attention Aggregation Model]{
        \centering
        \includegraphics[width=0.22\textwidth]{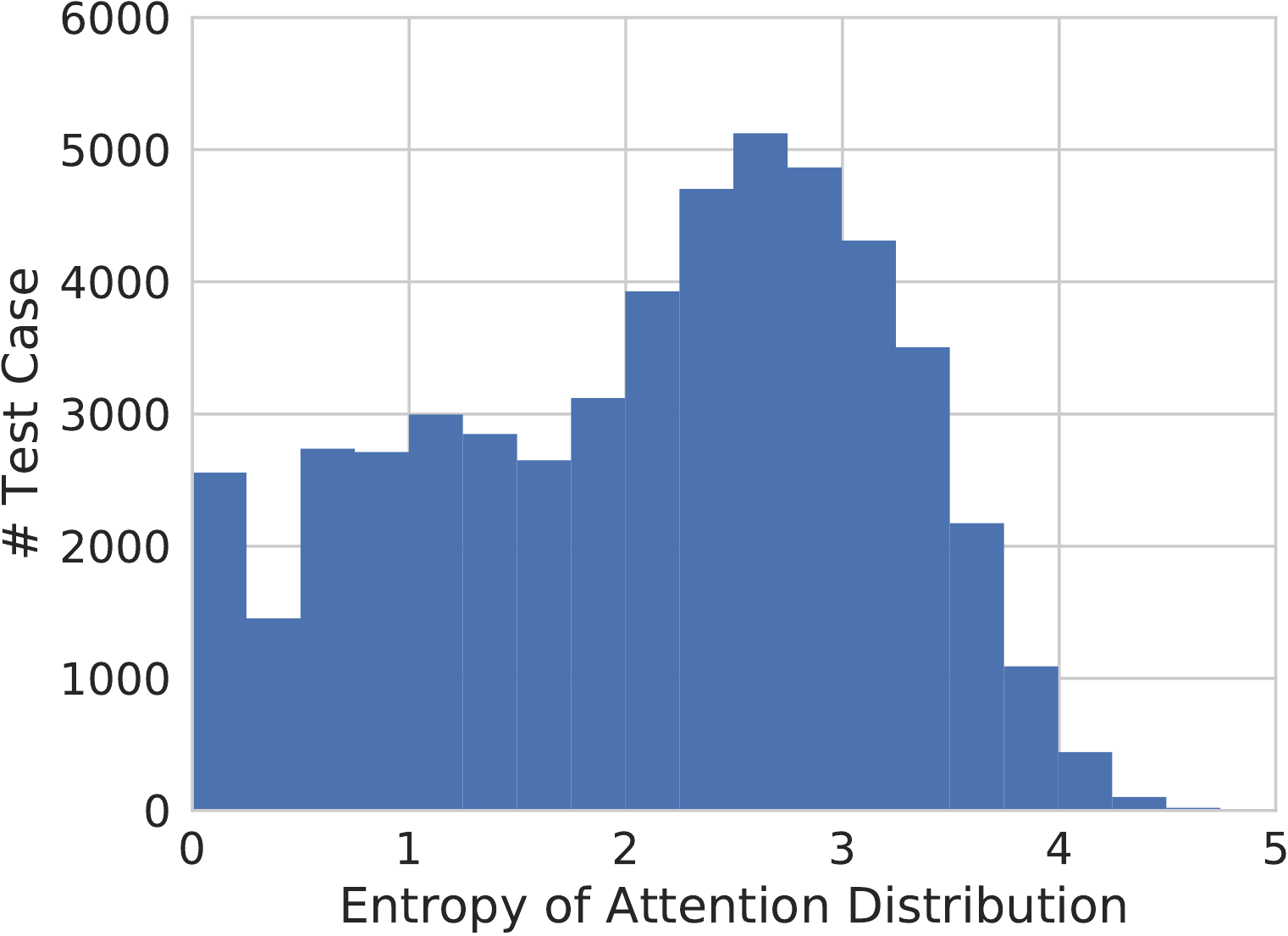}
    }
    \subfigure[Neighborhood Interaction Model]{
        \centering
        \includegraphics[width=0.22\textwidth]{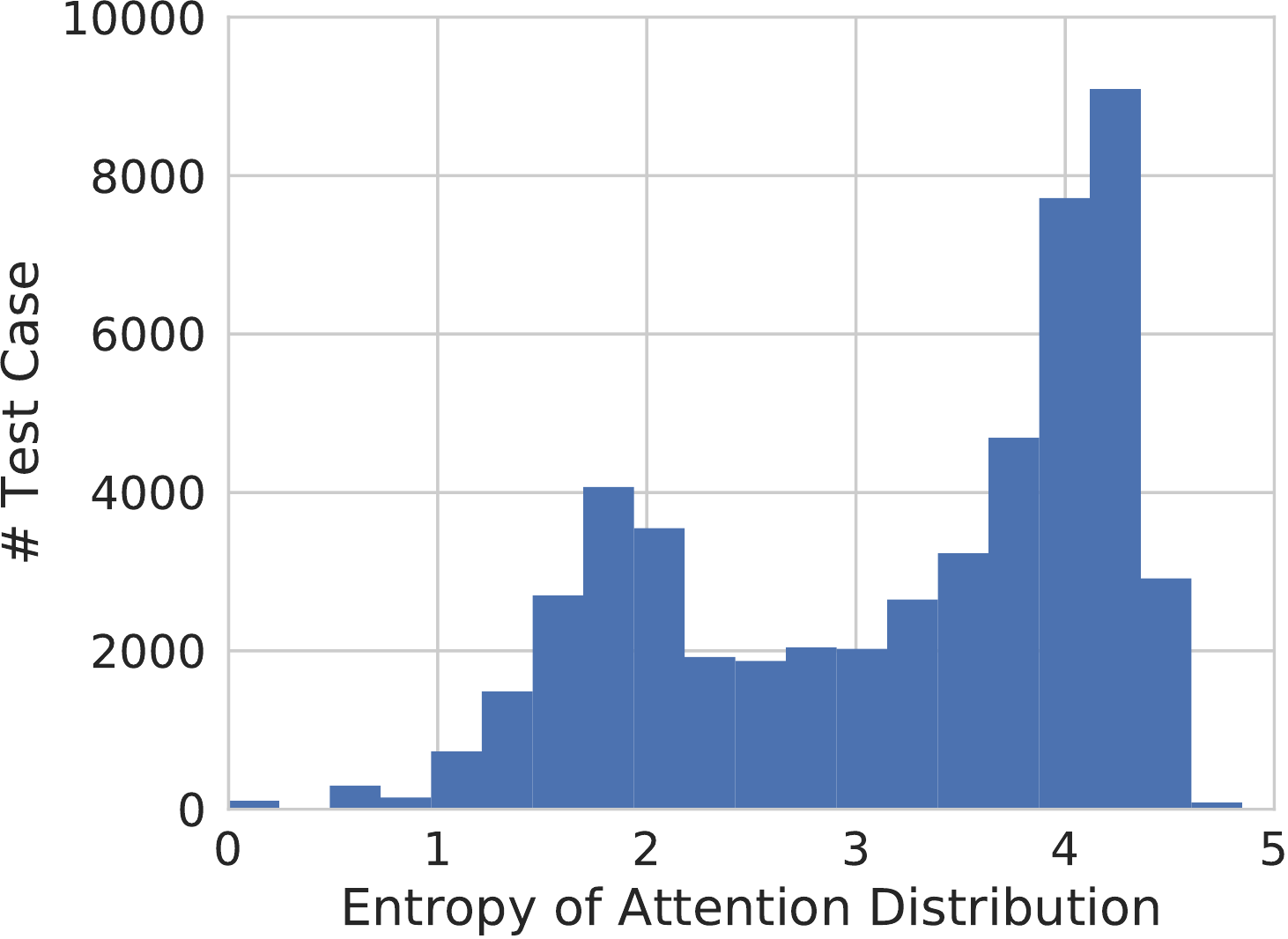}
    }
    \caption{Entropy histogram. \emph{Note:} The x-axis represents the entropy of attention distribution.}
    \label{fig:hist}
\end{figure}

To show the early summarization problem discussed in Section \ref{sec:ni}, as well as to understand how NI model improves other models, 
we conduct a case study on in this section. We randomly choose 10k users, 6k items, and 250k responses from the MovieLens-20M dataset, and randomly split training/validation/test sets at 6:2:2. 
We compare attention aggregation model (AAM) (Eq.\eqref{eq:atn}) and NI (Eq.~\eqref{eq:ni}) solely on the user-item interaction graph. Recall the general form of graph-based recommendation models in Eq.~\eqref{eq:form}, \ie $\hat{y} = \sigma(\mathbf{A} \odot \mathbf{Z})$. 
AAM learns the weight matrix $\mathbf{A}$ through user-side and item-side attention networks separately, \ie $\mathbf{A}_{i,j} = \alpha_{u,i} \alpha_{v,j}$, yet NI learns from both sides, \ie $\mathbf{A}_{i,j} = \alpha_{i,j}$. 

Since the elements in $\mathbf{A}$ sum to 1, a weight matrix can be regarded as a distribution. Thus we can calculate its entropy to quantitatively measure the information it contains. We calculate the entropy of the weight matrix $\mathbf{A}$ of each test sample and plot the histograms of entropy in Fig.~\ref{fig:hist}. The x-axis represents the entropy value, the larger value it has, the more information it contains. 
We can see that the weight matrices in NI model have higher entropy, \ie more informative. Besides, the average entropy of GAT is 2.12, and 3.18 for NI. Considering the significant improvements of NI over RippleNet (a special case of AAM) in Table~\ref{tab:ctr}, these results confirm the early summarization problem, and our NI model has the capability to learn more informative neighborhood interactions.

\begin{figure}
    \subfigure[$\mathbf{A}_{i,j}$ of AAM]{
        \centering
        \includegraphics[width=0.22\textwidth]{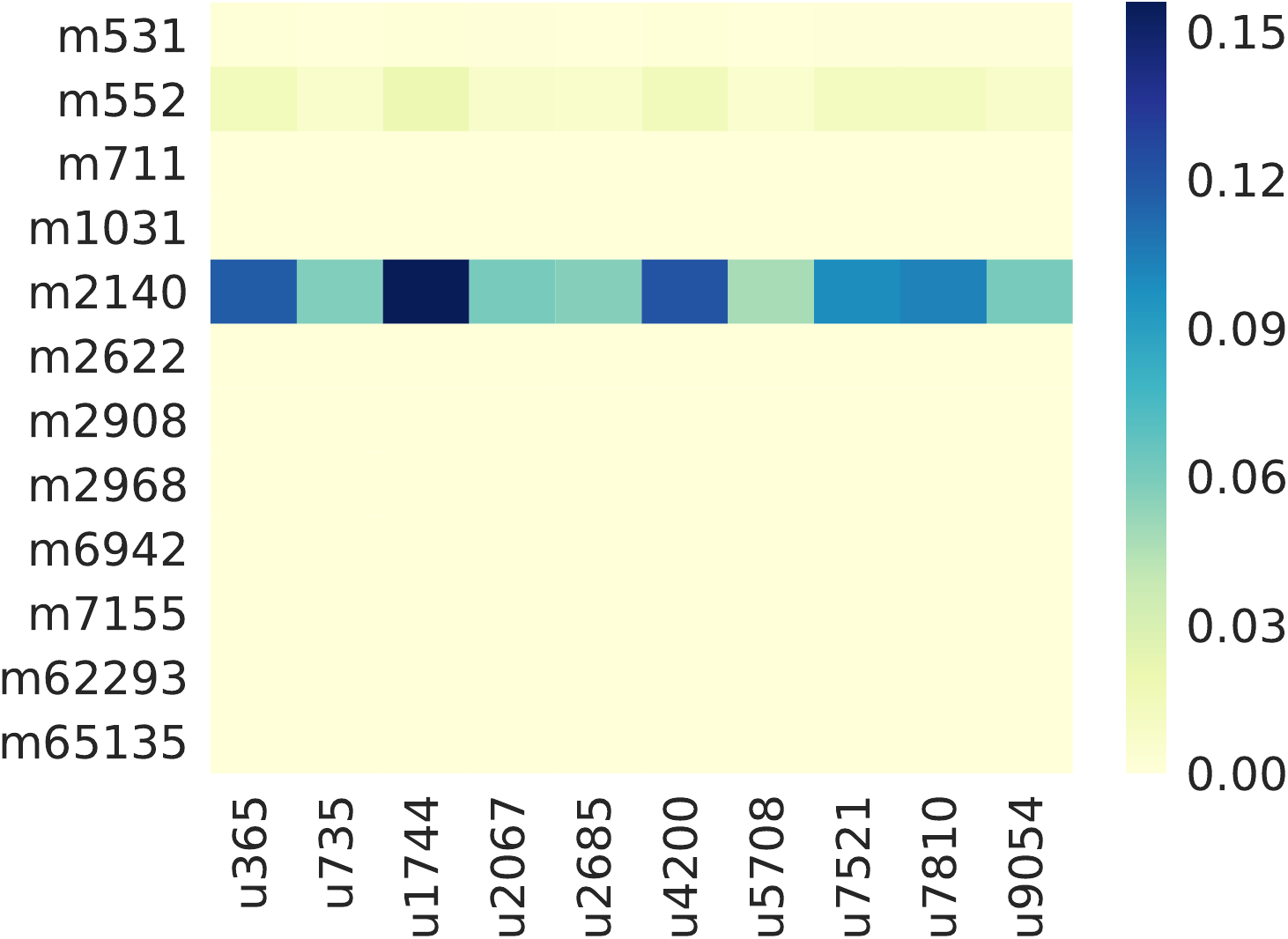}
    }
    \subfigure[$\mathbf{Z}_{i,j}$ of AAM]{
        \centering
        \includegraphics[width=0.22\textwidth]{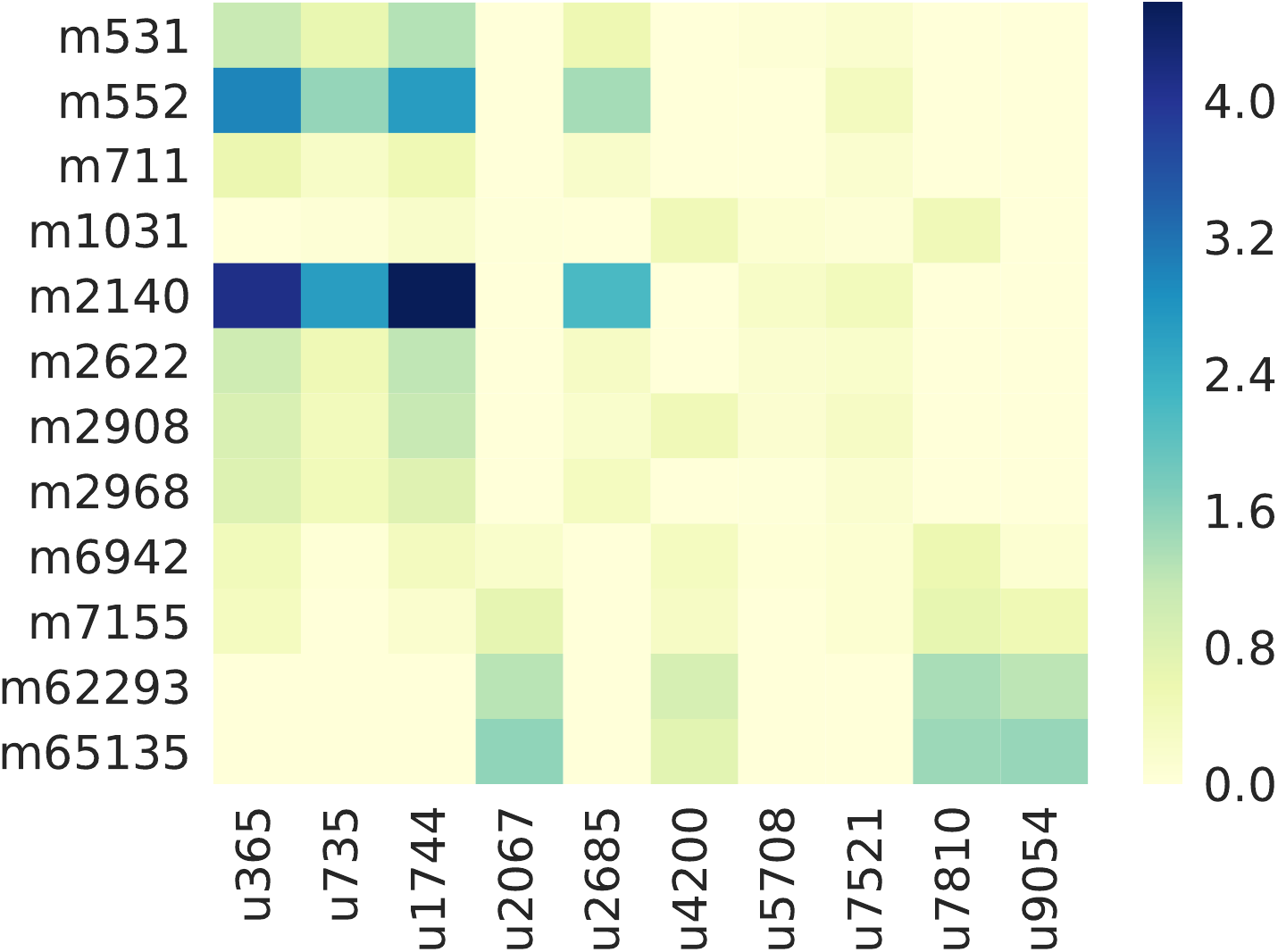}
    }
    \subfigure[$\mathbf{A}_{i,j}$ of NI]{
        \centering
        \includegraphics[width=0.22\textwidth]{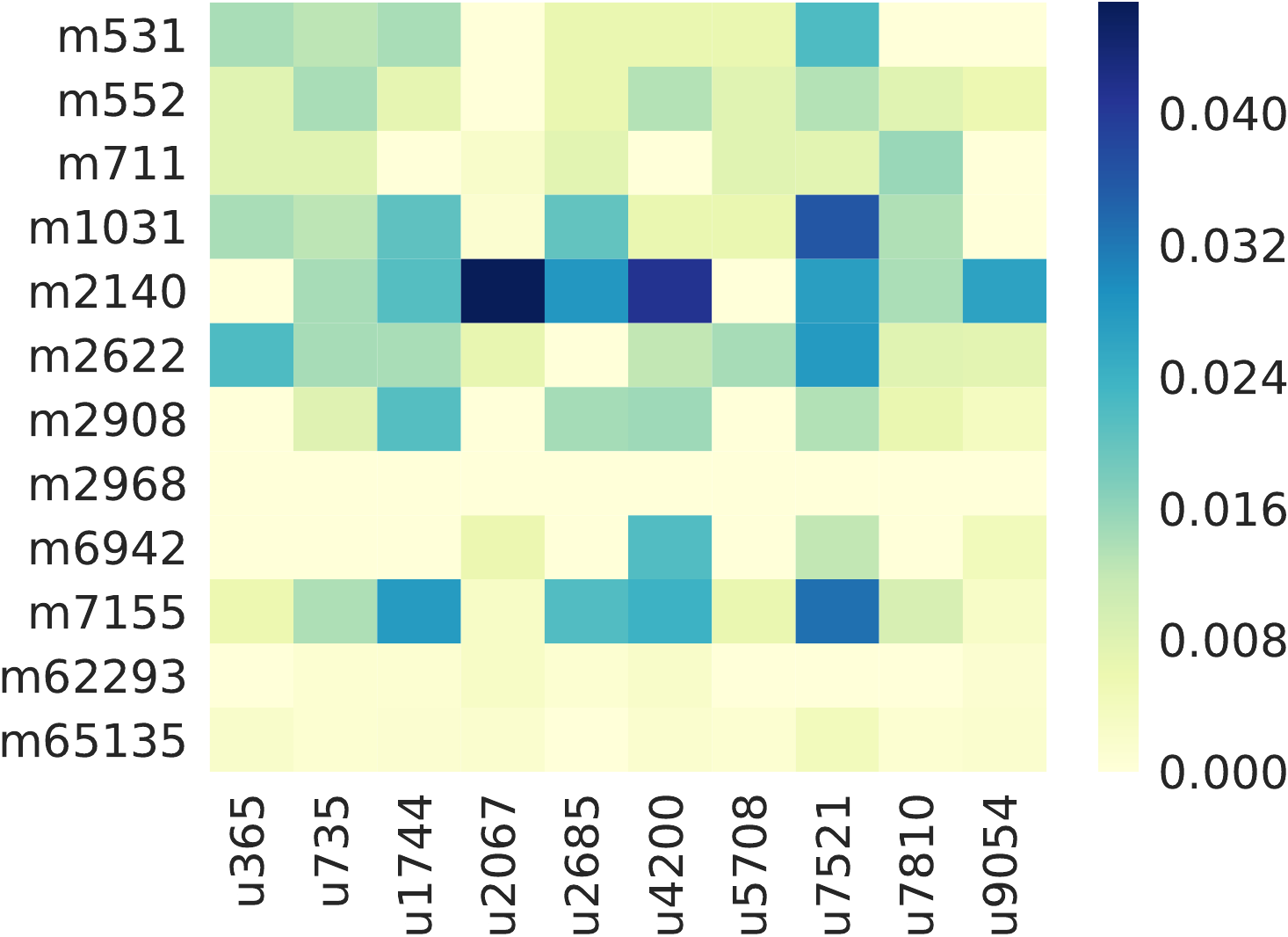}
    }
    \subfigure[$\mathbf{Z}_{i,j}$ of NI]{
        \centering
        \includegraphics[width=0.22\textwidth]{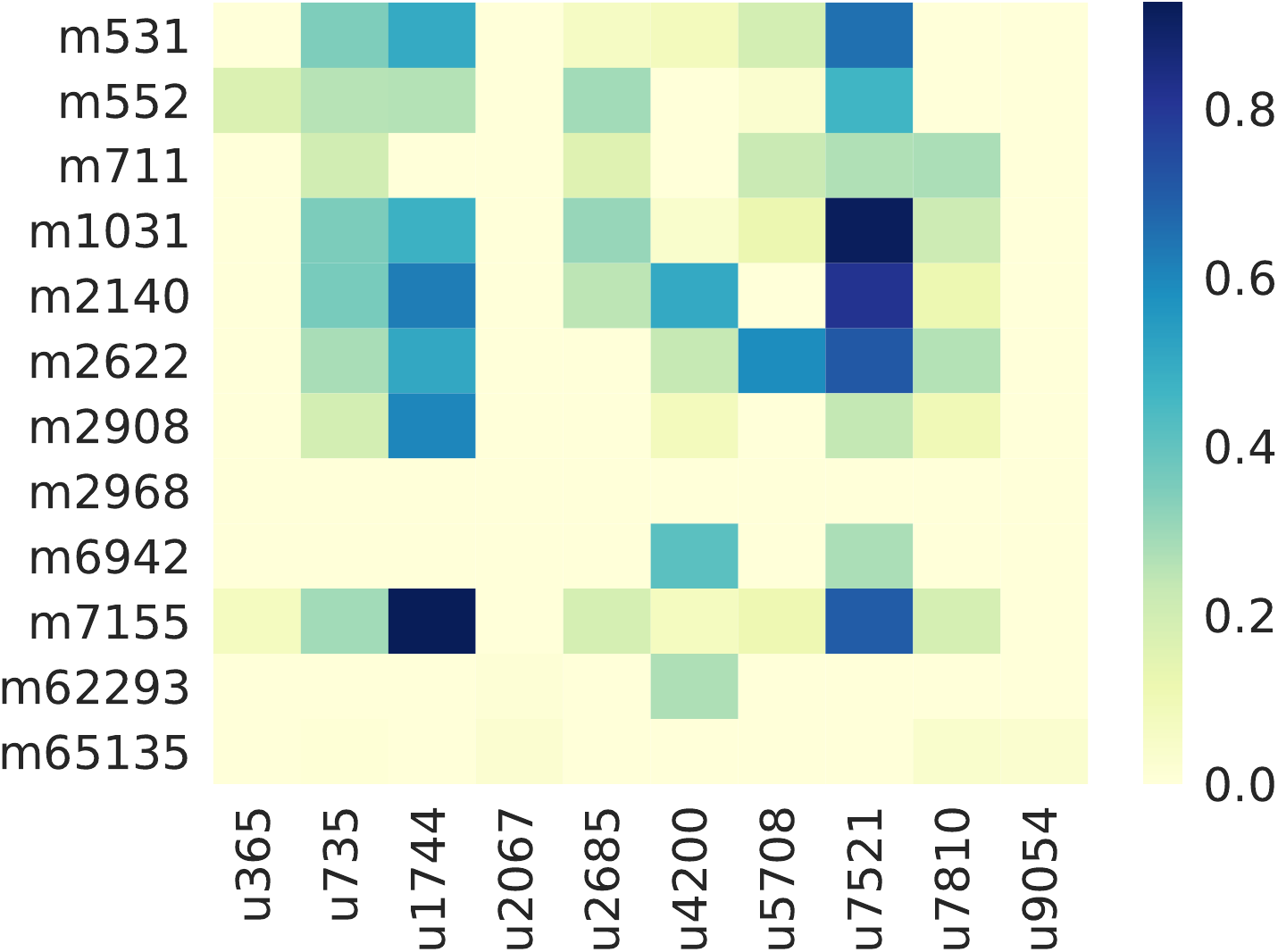}
    }
    \caption{Case study of test case (u46, m3993). \emph{Note}: In (a)-(d), the y-axis represents neighbors of user 46, and the x-axis represents neighbors of item 3993. AAM: attention aggregation model (Eq.~\eqref{eq:atn}).}
    \label{fig:case}
\end{figure}

We also randomly select a user-item pair (``u46'', ``m3993'') from the test set and plots the weight matrix $\mathbf{A}$ and interaction matrix $\mathbf{Z}$. We compare AAM and NI in Fig.~\ref{fig:case}. The x-axis represents the neighbors of item ``m3993'', and y-axis for the neighbors of user ``u46''. In user-item interaction graph, users are linked to items with positive feedbacks. Thus user neighbors are items, and item neighbors are users. Grids with darker colors have larger values.

We can observe that:
(i) Comparing (a) and (c), we find AAM mainly focuses on a single neighbor ``m2140'' of the user, while NI focuses on many more other neighbor pairs. 
(ii) Comparing (a) and (b), we find AAM disregards those neighbor pairs with high interactions, \eg, (``m552'', ``u1744''). While in (c) and (d), we find NI preserves more neighbor pairs with high interactions.
(iii) Checking in training set, we find the pairs with high interactions in our NI model, such as (``m7155'', ``u1744''), (``m1031'', ``u7521'') and (``m2140'', ``u1744'') are positive samples, which should be fully considered in prediction. 
Based on the above observations, we conclude AAM may lose useful information after compressing neighborhood information into single representation, while NI can preserve more useful information. %Not limited to user/item representations, NI can dynamically incorporate different neighbors into recommendation.

\section{Related Work} \label{sec:re}
Our work is highly related with knowledge-enhanced recommendation, and graph representation models. 

\subsection{Knowledge-enhanced Recommendation}

Traditional recommender systems 
mostly suffer from several inherent issues such as data sparsity and cold start problems. 
To address the above problems, researchers usually incorporate side information. 
The utilization of side information mainly categorizes into 3 groups.

The first is feature-based, which regards side information as plain features and concatenates those features with user/item IDs as model input, including Matrix factorization models~\cite{koren2008factorization, juan2016field}, DNN models~\cite{qu2016product, guo2017deepfm, qu2018product}, etc. Feature-based models highly rely on manual feature engineering to extract structural information, which is not end-to-end and less efficient. 

The second way is meta path-based, which builds heterogeneous information network (HIN) on the side information. 
For example, PER~\cite{yu2014personalized} and FMG~\cite{zhao2017meta} extract meta path/meta graph-based features to represent the connection between users and items along different types of relation paths.
MCRec~\cite{hu2018leveraging} instead learns context representations from meta paths to facilitate recommendation. 
DeepCoevolve~\cite{dai2016deep} further leverages user-item ineteraction network in sequential recommendation. 
Though these models are more intuitive, they usually require much expertise in meta-path design, making them less applicable in scenarios with complex schema. 

Compared with the previous 2 ways, external knowledge graph contains much more fruitful facts and connections about items~\cite{bollacker2008freebase}.
For example, CKE~\cite{zhang2016collaborative} proposes a general framework to jointly learn from the auxiliary knowledge graph, textual and visual information. 
DKN~\cite{wang2018dkn} is later proposed to incorporate knowledge embedding and text embedding for news recommendation. 
More recently, RippleNet~\cite{wang2018ripplenet} is proposed to simulate user preferences over the set of knowledge entities. It automatically extends user preference along links in the knowledge graph, and achieves state-of-the-art performance in knowledge graph-based recommendation. 
The major difference between prior work and ours is that NI focuses more on the interactions between neighbor nodes, and predict from graph structures directly.

\subsection{Graph Representation}
Graph representation learning aims to learn latent, low-dimensional representations of graph vertices, while preserving graph topology structure, node content, and other information.
In general, there are two main types of graph representation methods: unsupervised and semi-supervised methods. 

Most of the unsupervised graph representation algorithms focus on preserving graph structure for learning node representations~\cite{perozzi2014deepwalk, grover2016node2vec, tang2015line}. For example, DeepWalk~\cite{perozzi2014deepwalk} uses random walks to generate node sequences and learn node representations. Node2vec~\cite{grover2016node2vec} further exploits a biased random walk strategy to capture more flexible contextual structures. LINE~\cite{tang2015line} uses first-order and second-order proximity to model a joint probability distribution and a conditional probability distribution on connected vertices. 

Another type is semi-supervised models~\cite{huang2017label, kipf2016semi, velickovic2017graph}. In this type, there exist some labeled vertices for representation learning. 
For example, 
LANE~\cite{huang2017label} incorporates label information into the attributed network embedding while preserving their correlations.
GCN~\cite{kipf2016semi} utilizes a localized graph convolutions for a classification task. 
GAT \cite{velickovic2017graph} uses self-attention network for information propagation, which utilizes a multi-head attention mechanism to increase model capacity.
GCN and GAT are popular architectures of the general graph networks, and can be naturally regarded as plug-in graph representation modules in other supervised tasks.
In this work, we mainly utilize graph networks to generate structural node embeddings for KIG.

\section{Conclusion} \label{sec:co}

In this paper, we review previous graph-based recommender systems and find an early summarization problem of the existing methods. 
We extend user-item interactions to neighbor-neighbor interactions, and propose Neighborhood Interaction (NI) to further explore the neighborhood structures of users and items.
Integrating high-order neighborhood information with Graph Neural Networks and Knowledge Graphs into NI, we obtain an end-to-end model, namely Knowledge-enhanced Neighborhood Interaction (KNI). We compare KNI with state-of-the-art models on 4 real-world datasets, and the superior results of KNI on CTR prediction and top-N recommendation demonstrate its effectiveness. We also provide a case study to quantitatively measure the early summarization problem. In the future, a promising direction is extending neighborhood interactions to higher-orders. Another direction is integrating user-side information in KIG to adapt to more general scenarios.

\begin{acks}
We would like to thank the support of National Natural Science Foundation of China (61632017, 61702327, 61772333), Shanghai Sailing Program (17YF1428200). Jian Tang is supported by the Natural Sciences and Engineering Research Council of Canada, as well as the Canada CIFAR AI Chair Program.
\end{acks}

\bibliographystyle{ACM-Reference-Format}
\bibliography{RS-KG}

\end{document}